# Nanoscale Non-Destructive Ferroelectric Characterization with Non-Contact Heterodyne Electrostrain Force Microscopy

Qibin Zeng,[1] Qicheng Huang,[2] Hongli Wang,[1,5] Caiwen Li,[2,3] Zhen Fan,[2] Deyang Chen,[2,3] Yuan Cheng,[4,6] Kaiyang Zeng[1,7]*

[1]Department of Mechanical Engineering, National University of Singapore, Singapore 117576, Singapore

[2]Institute for Advanced Materials, South China Academy of Advanced Optoelectronics, South China Normal University, Guangzhou 510006, China

[3]Guangdong Provincial Key Laboratory of Optical Information Materials and Technology, South China Academy of Advanced Optoelectronics, South China Normal University, Guangzhou 510006, China

[4]Institute of High-Performance Computing, Agency for Science Technology and Research, Singapore 138632, Singapore

[5]The Key Lab of Guangdong for Modern Surface Engineering Technology, National Engineering Laboratory for Modern Materials Surface Engineering Technology, Institute of New Materials, Guangdong Academy of Sciences, Guangzhou, 510650, China

[6]Monash Suzhou Research Institute, Suzhou 215123, China

[7]NUS (Suzhou) Research Institute (NUSRI), Suzhou 215123, China

*Corresponding author: Kaiyang Zeng

**Email:** mpezk@nus.edu.sg

**ORCID:** Kaiyang Zeng 0000-0002-3348-0018

**Abstract**

Perceiving nanoscale ferroelectric phenomena from real space is of great importance for elucidating underlying ferroelectric physics. During the past decades, nanoscale ferroelectric characterization has mainly relied on the Piezoresponse Force Microscopy (PFM), however, the fundamental limitations of PFM have made the nanoscale ferroelectric studies encounter significant bottlenecks. In this study, a high-resolution non-contact ferroelectric measurement, named *Non-Contact Heterodyne Electrostrain Force Microscopy* (*NC-HEsFM*), has been introduced *firstly*. It has been unambiguously demonstrated that NC-HEsFM can operate on multiple eigenmodes to perform ideal high-resolution ferroelectric domain mapping, standard ferroelectric hysteresis loop measurement and controllable domain manipulation. With using quartz tuning fork (QTF) sensor and heterodyne detection, NC-HEsFM shows an unprecedented capability in achieving real non-contact yet non-destructive ferroelectric characterization with negligible electrostatic force effect. It is believed that NC-HEsFM can be extensively used in various ferroelectric or piezoelectric studies with providing substantially improved characterization performance. Meanwhile, the QTF-based force detection makes NC-HEsFM highly compatible for high-vacuum and low-temperature environments, providing ideal conditions for achieving an ultra-high spatial resolution to investigate the most intrinsic ferroelectric phenomena.

**Introduction**

Ferroelectric materials underpin a broad variety of modern electronic devices, such as ultrasonic sensors, actuators, transducers, ferroelectric memories and photovoltaic cells (1-5). Understanding the fundamental mechanisms of ferroelectric phenomena is of great importance in both science and applications. Invented in 1992, the Piezoresponse Force Microscopy (PFM) was firstly introduced as a powerful local probing technique to study the properties and phenomena of the ferroelectrics, bringing researchers a ground-breaking opportunity to perceive the nanoscale ferroelectricity from real space (6). Through capturing the piezoelectric strain information at the nanoscale, PFM unambiguously provides an acute insight for understanding those fundamental physics of the ferroelectric and piezoelectric materials (3, 7-11). After the inception of nearly 30 years, PFM has successfully been developed as a versatile yet indispensable tool for the study of ferroelectricity, piezoelectricity as well as other electromechanical coupling phenomena (3, 7, 12-18). However, the extensive applications and studies of the PFM have revealed a growing number of challenges and concerns about this important technique, which is greatly challenging its validity in many studies for numbers of years (7, 19, 20). It has been widely recognized that the two most significant issues faced by PFM are the contact-mode operation and the influences of the electrostatic force (12, 15). The contact-mode operation can induce pronounced modifications to both the sample and tip, such as the surface damage, contamination, charge injection and triboelectrification (15, 21-26) as well as the tip wear, contamination and damage (15, 23, 24, 26-29), which will all affect the PFM signal and may cause significant reproducibility problems and artefacts in the measurements. Since the sample and tip modifications are almost inevitable during the contact-mode PFM scanning, by far researchers can only have very limited solutions to reduce such effects, such as decreasing the setpoint of force feedback, using softer cantilever and mapping with a point-by-point manner (23, 30, 31). In addition to the modifications, the tip-sample contact force usually causes a significant stress field in the sample, which can potentially induce a remarkable flexoelectric polarization (32-34) and affect the ferroelectric polarization switching (35, 36). Furthermore, as there is a contact area formed at the tip-sample junction which contains a large number of interactional atoms, the contact-mode operation fundamentally limits the spatial resolution of PFM to the best of sub-10 nanometers (7, 15, 37), and a higher spatial resolution, such as the molecular or atomic resolution, can hardly be achieved in conventional PFM. Another important issue closely related to the contact-mode operation is the resonance tracking. As the piezoelectric strain induced by tip electric field is typically with the order of picometers, resonance amplification is usually necessitated in PFM to enhance the signal-to-noise ratio (SNR) and reduce the AC drive needed (8, 38-40). However, a significant concern with the resonance enhancement is that the resonance is highly sensitive to the tip-sample contact status, and a slightly change of the contact status may cause an apparent influence to the PFM signal (38). Therefore, once resonance enhancement is utilized in PFM, the resonance tracking issue is usually inevitable. Much efforts have been made by researchers to address the resonance tracking problem, and the proposed yet now widely used methods for resonance tracking in PFM are dual-frequency resonance-tracking (DFRT) and band-excitation (BE) techniques (41-43). Both DFRT and BE are based on the ideal simple-harmonic-oscillator (SHO) model to extract the resonance parameters, the difference is that DFRT uses two while BE uses multiple points of the frequency spectrum to fit the highly non-linear SHO model (30, 41-43). In consequence, any deviation from the ideal SHO model during the contact scanning process, especially for the DRFT method, will deteriorate the accuracy of the resonance tracking, and at the same time, the low SNR, especially for the BE method, can also induce large uncertainty to the tracking results.

On the other hand, PFM is based on inverse piezoelectric effect and thereby detects the piezoelectric deformation of the sample in response to an external electric field applied between the tip and sample, thus a concomitant electrostatic force will be always involved in the final PFM signal, causing large ambiguity to the interpretation of the measurement (8, 12, 15, 19, 20, 44, 45). Due to the great importance, a large amount of research work has been continuously implemented to eliminate or quantify the electrostatic force contribution in the PFM measurements (3, 8, 12, 15, 20, 22, 46-57), but nearly all of the proposed solutions, including using stiff cantilever, applying DC compensation voltage, imaging in liquid environment, using higher eigenmode and interferometric detection as well as off-line analysis strategies, are subject to many specific limitations (12, 48, 51, 54-60). Until very recently, the inception of Heterodyne Megasonic Piezoresponse Force Microscopy (HM-PFM) provides a much more effective solution to eliminate the influence of both local and distributed electrostatic forces by using high-frequency excitation and heterodyne detection scheme (61, 62). The HM-PFM, however, is still based on the contact-mode operation.

Although the fundamental contact-mode and electrostatic force issues have affected the PFM for several decades, due to the intrinsic configuration and principle of PFM, by far the effective solutions are still very limited especially when facing the contact-mode issue, and solving these two intractable problems simultaneously are even more difficult. Given these technical limitations of PFM, it has been gradually recognized that to achieve an ideal piezoelectric strain measurement using conventional PFM could be a great challenge. Therefore, exploiting a brand-new scanning probe technique which provides similar function of PFM but has much better performance to substitute the role of PFM becomes more and more significant. However, this type of study is highly challengeable and to date such substitutions have rarely been proposed. In this study, an excellent candidate for substituting the conventional PFM, which is called Non-Contact Heterodyne Electrostrain Force Microscopy (NC-HEsFM) (63), has been developed and introduced for the first time. Similar to the PFM, NC-HEsFM also measures the tip electric field-induced surface strain, but *via* using the brand-new quartz tuning fork (QTF) force sensor and heterodyne detection scheme, a true non-contact yet non-destructive surface strain measurement with significantly minimized electrostatic force effect have been successfully achieved in a prototype of NC-HEsFM. As it has been generally accepted nowadays that the electric field-induced surface deformation in PFM can originate from multiple mechanisms (19, 20), the terminology "piezoresponse" in PFM, coined by Gruverman et al. in 1995 (19, 64), may indeed cause misunderstandings when using PFM to study many non-piezoelectric phenomena, a typical example is that the PFM has been renamed as Electrochemical Strain Microscopy when electrochemical Vegard strain is measured (65). Herein, to avoid such ambiguity, we define any surface strain (or displacement) induced by an external electric field as "electrostrain". Within this definition, the electrostrain in PFM or NC-HEsFM measurements may arise through a variety of mechanisms including piezoelectricity (66), electrochemical Vegard strain (65), electrostriction (67), measurement-induced electromechanical coupling (due to flexoelectric polarization or tip electric field-induced electrochemical dipole) (33, 68), converse flexoelectricity (69), Maxwell stress (70), thermal expansion (due to Joule heating) (71), and perhaps a mixture of more than one of these effects. In this study, we are focusing on the most important application of NC-HEsFM, i.e., the measurement of piezoelectric strain on ferroelectric materials. In this paper, the principle and technical design towards realizing non-contact detection of electrostrain has been discussed firstly. Then several conventional ferroelectric materials are used to test the basic ferroelectric domain characterization and manipulation, switching spectroscopy and multi-eigenmode operation capabilities of the NC-HEsFM. The results unambiguously show that NC-HEsFM can operate on multiple eigenmodes to perform ideal high-resolution ferroelectric domain mapping, standard ferroelectric hysteresis loop measurement and artificial domain writing with a true non-contact manner. At the same time, the superior non-contact yet real non-destructive characterization and electrostatic force minimization capabilities of the NC-HEsFM have been demonstrated, which are compared with the results from conventional PFM. Finally, the mechanism of achieving effective electrostrain detection by NC-HEsFM has been analyzed using theoretical model and finite element simulations.

## Results and Discussion

**Non-contact detection of electrostrain.** Typically, the tip electric field-induced electrostrain has a surface displacement at picometer level, thereby directly probe this tiny displacement within such a small tip-sample interaction region is almost impractical. Conventional PFM utilizes an indirect manner to achieve the detection of electrostrain, in which is to transfer the dynamic displacement of the surface to the cantilever motion via local tip-sample interaction. Then the cantilever motion can be easily measured by monitoring its deflection using an optical lever system. As the tip-sample interaction significantly decays with increasing tip-sample separation, to provide a strong enough yet stable tip-sample interaction, PFM is always operated at the static contact mode to obtain the best performance. However, as mentioned above, this contact-mode operation does become a fundamental limitation of the PFM. To overcome this contact-mode issue, a natural yet effective solution is to operate the PFM with a non-contact mode, but this "simple" idea does become a long-standing challenge in PFM research. To date, only a few attempts of operating PFM with an intermittent contact mode have been demonstrated, whereas the results do show a significant deterioration of SNR and sensitivity as well as a stronger influence from the electrostatic force (24, 72). Alternatively, conducting PFM measurement with a point-by-point manner has also been proposed to reduce the duration of tip-sample contact (30, 31), however the associated issues of complicated scanning control and much longer imaging time have limited the utilization of this method, and in fact, the electrostrain measurement at each point is still performed *via* the conventional contact mode. The main difficulty towards achieving non-contact electrostrain detection originates from the significant attenuation of both the electric field located around the tip-sample junction and the tip-sample interaction force with increasing tip-sample distance (will be discussed latter). The significant decrease of the electric field strength minimizes the stimulated sample electrostrain while the decay of the tip-sample interaction causes the electrostrain detection sensitivity decreased pronouncedly, which in turn makes the electrostatic force effect being largely enhanced. Therefore, to achieve an effective non-contact electrostrain measurement, the tip apex should keep a distance as small as possible from the sample surface, and if the tip is oscillating, the oscillation amplitude should also be as small as possible. However, the theoretical analysis and experiments have clearly demonstrated that, when using conventional cantilever-based tip to perform non-contact scanning at the very near surface, a minimum amplitude of the cantilever is required to prevent the occurrence of the "jump-to-contact" phenomenon (73, 74). This minimum amplitude increases when soft cantilever is used, e.g., for a cantilever with medium stiffness of 17 N/m, the minimum amplitude is found to be ~34 nm at the very near surface (75, 76), which is a too large tip-sample separation for effective detection of electrostrain (will be discussed later). Consequently, it would be of great difficulty to achieve a true non-contact detection for electrostrain if the conventional cantilever sensor and optical detection system are used.

Since the classic cantilever sensor suffers from fundamental limitations towards achieving non-contact electrostrain measurement, it brings a question that whether is possible to replace this most widely used cantilever sensor by other types of force sensors. The answer is yes, and the QTF sensor is exactly the ideal and powerful candidate. QTF was firstly introduced to Non-contact Atomic Force Microscopy (NC-AFM) as the atomic force sensor during the end of 1990s (77-79), and thereafter, the QTF-based NC-AFM has received intensive attentions and has been extensively applied to study the most frontier topics of surface science, especially for those systems under extreme conditions (i.e., ultra-high vacuum and low temperature) where ultra-high spatial resolution can be achieved (76, 80-83). The most important advantages of QTF sensor are its ultra-high stiffness, piezoelectric self-detection capability and high resonance quality factor ($Q$ factor) (76, 78). The ultra-high stiffness of the QTF prong can totally eliminate the jump-to-contact effect thus allowing the tip to oscillate at the position very near the sample surface even with an ultra-low amplitude (76, 84). Meanwhile, the piezoelectric self-detection capability and high resonance $Q$ factor can perfectly provide a much simpler yet high-sensitive force detection scheme, which greatly supports the operation of QTF in a variety of environments, including ultra-high vacuum, ambient and liquid phase (80, 85, 86), with an ultra-low vibration amplitude. Hence, based on these facts and advantages, it is conspicuous that QTF is almost an ideal force sensor which can be used to reach the long-awaited goal of non-contact electrostrain measurement, as it can avoid almost all the fundamental limitations suffered by the classic cantilever force sensor. Meanwhile, refer to the ultra-high spatial resolution achieved in QTF-based NC-AFM (76, 80), it is reasonable to believe that the current resolution limit of PFM measurement may get a breakthrough by using QTF sensor under appropriate conditions/environments. Since the traditional cantilever sensor requires a complex optical detection system, using conventional PFM to study piezo/ferro-electricity under extreme conditions becomes very difficult. In contrast, the piezoelectric self-detection and small volume of QTF sensor make it highly compatible for extreme environments (76, 82), implying that to study the fundamental physics of piezo/ferro-electricity under these environments will be much easier if using the QTF sensor. As the QTF-based NC-AFM also perfectly supports the AFM/Scanning Tunneling Microscopy (STM) dual-mode (76, 87), when using QTF sensor to measure the electrostrain on ultra-thin films (88), the tunneling current or tunneling spectroscopy can be measured *in-situ* or even simultaneously thus more abundant physical information can be attained. In addition, it has been reported that the electrostrain measurement by conventional PFM can be significantly affected by the Joule heating associated thermal expansion (71). However, if QTF sensor is used for non-contact scanning, the gap between the sample surface and tip apex can help to reduce the tip-sample current, thereby the influence of Joule heating on electrostrain measurement can be minimized. Based on the multiple advantages discussed above, the QTF attached with a conductive nano-tip is introduced here, for the first time, as the force sensor to simultaneously measure the topography and electrostrain with a true non-contact manner (63).

**Multi-frequency operation and heterodyne detection.** To achieve the force measurement and feedback control, the QTF needs to be operated in the dynamic mode where a flexural mode of the QTF is typically excited and the corresponding resonance parameters, i.e., the resonance amplitude, phase and frequency, are monitored as the control signal. For most of the previous studies, the 1st flexural mode of the QTF is utilized to measure topography as this mode has better force sensitivity (77, 78, 89), while several studies have demonstrated that the 2nd flexural mode can also provide similar topography measurement capability (90, 91). In this work, considering the very tiny electrostrain and ~40 times higher effective stiffness of the 2nd flexural mode (90), to optimize the sensitivity, the 1st flexural mode is employed to detect the electrostrain while the 2nd flexural mode, operated in a classic frequency modulation mode (92), is used for topography measurement. As the electrostrain and topography measurements are completed by two flexural modes independently, it seems that the same technique principle of PFM can be directly applied to this new system. However, as the QTF is a capacitor-like device without any electric shielding, a severe capacitance cross-talk will be induced in the QTF when AC drive is applied to the sample, causing a very strong background interference in the output of the QTF (Supplementary Materials, S1). Therefore, if the conventional PFM's homodyne detection is used, the target electrostrain signal will be totally masked by this strong background interference, causing the demodulation of the electrostrain signal impossible. In addition, since there is no electric shielding, the QTF will also be excited by the sample's AC drive as well as the associated electrostatic force, which will introduce an apparent parasitic oscillation and then make the measured electrostrain signal largely ambiguous. Therefore, the classic homodyne detection scheme used in conventional PFM cannot be transplanted to the QTF-based system to measure electrostrain. Recently, the inception of HM-PFM provides a remarkably different way to measure the electrostrain by using the heterodyne detection method (61, 62). In HM-PFM, the frequencies of the target electrostrain signal and the sample's AC drive are largely different, which tactfully makes the co-frequency interferences in the conventional PFM, typically the electrostatic force, can be significantly minimized. Based on this novel design, the above-mentioned co-frequency interferences in QTF system, including the capacitance cross-talk and parasitic oscillation, can also be perfectly avoided if heterodyne detection scheme is adopted.

To realize heterodyne detection for the dynamic electrostrain, another oscillation component should be provided as the reference, then the electrostrain and the reference oscillations can be mechanically mixed to obtain the difference-frequency component through the non-linear tip-sample interaction (61, 62). Here, a higher $n^{th}$ ($n \geq 3$) flexural mode of the QTF is excited mechanically and the correspondingly induced vibration at the tip apex is $A_h \sin(2\pi f_h t + \phi_h)$ (where $A_h$, $f_h$, $t$ and $\phi_h$ are amplitude, frequency, time and phase of the tip vibration, respectively), which is used as the reference oscillation for heterodyne detection. Considering the sample's electrostrain vibration as $A_s\sin(2\pi f_s t + \phi_s)$ (where $A_s$, $f_s$ and $\phi_s$ are amplitude, frequency and phase of the sample vibration, respectively), the time-dependent tip-sample interaction force (when tip is at the near surface) is given by:

$$F_{ts}(z)=F_{ts}\left[z_0 + A_h \sin(2\pi f_h t + \phi_h) - A_s \sin(2\pi f_s t + \phi_s)\right] \qquad (1)$$

where $z$ and $z_0$ are the instantaneous and equilibrium tip-sample separations, respectively. Since the tip-sample interaction force $F_{ts}(z)$ depends non-linearly with $z$, under the condition of small vibration amplitudes, $F_{ts}(z)$ can be approximately expressed by a Taylor series at $z = z_0$ up to second order:

$$F_{ts}(z) \approx F_{ts}(z_0) + F'_{ts}(z_0)(z-z_0) + \frac{1}{2}F''_{ts}(z_0)(z-z_0)^2 \qquad (2)$$

Combining Equation (1) and (2) and then ignoring all the static and high-frequency items, the difference-frequency force item of interest is:

$$F_{ts}(z)_{diff} = -\frac{1}{2}F''_{ts}(z_0)A_h A_s \cos\left(2\pi f_{diff} t + \phi_s - \phi_h\right) \qquad (3)$$

in which $f_{diff} = f_s - f_h$ is the difference frequency. The difference-frequency force $F_{ts}(z)_{diff}$ is the product of the mechanical frequency mixing process, which contains the electrostrain information of interest. It is obvious that any change of the amplitude $A_s$ and phase $\phi_s$, such as $A_s$ decreasing at the ferroelectric domain wall and 180° $\phi_s$ change between the oppositely polarized domains, will be reflected by the changes of $F_{ts}(z)_{diff}$. To detect this difference-frequency force by QTF, $f_{diff}$ is intentionally set to be the resonance frequency of the 1$^{st}$ flexural mode thus resonance enhancement can be used to obtain a better detection sensitivity and SNR. From Equation (3), it is clear that if the 2$^{nd}$-order force gradient $F''_{ts}(z_0)$, the amplitude $A_h$ and phase $\phi_h$ of the $n^{th}$ flexural mode are constant, the amplitude and phase of the electrostrain vibration can be extracted from the motion of the 1$^{st}$ flexural mode. However, as the resonances of both the 1$^{st}$ and $n^{th}$ mode is employed to obtain the difference-frequency vibration (refer to as difference-frequency electrostrain, DFE, hereafter), the resonances of these two modes should keep being tracked to avoid the influences of resonance shift (i.e., facing the resonance tracking issue). Fortunately, the frequency shift-based height feedback regime in NC-HEsFM does provide the capability to maintain a constant $F''_{ts}(z_0)$ while track the resonances of both the 1$^{st}$ and the $n^{th}$ flexural mode. In the NC-HEsFM, the resonance frequency shift of the 2$^{nd}$ flexural mode ($\Delta f_2$) is used as the height feedback signal, and during the non-contact scanning process, the feedback control keeps adjusting the tip height to maintain $\Delta f_2$ constant at the setpoint. As the amplitude of the 2$^{nd}$ flexural mode is usually quite small, $\Delta f_2$ can be approximately proportional to the 1$^{st}$-order force gradient $F'_{ts}(z_0)$ (i.e., tip-sample contact stiffness) (73) and thus a constant $\Delta f_2$ will indicate a constant $F'_{ts}(z_0)$ (90). Therefore, both the resonances of the 1$^{st}$ and $n^{th}$ flexural mode can be tracked automatically during the NC-HEsFM scanning since the resonance frequency of each mode is determined by $F'_{ts}(z_0)$ (93). Comparing with the previous DRFT and BE techniques used in the conventional PFM, the frequency shift-based height feedback unambiguously provides a much more straightforward yet accurate way to achieve resonance tracking. Typically, the height feedback makes the tip located within the monotone interval of $F'_{ts}(z)$, according to the basic Lennard-Jones potential model (73), $F''_{ts}(z)$ is also monotonic within this monotone interval of $F'_{ts}(z)$ (see Supplementary Materials, Fig. S8), thus a constant $F'_{ts}(z_0)$ will indicate a constant $F''_{ts}(z_0)$ if the tip-sample interaction potential keeps unchanged. As the tip-sample interaction potential is dominated by the atoms of the tip and sample, if the sample component is uniform within the scanning area, the constant $F''_{ts}(z_0)$ condition can be reached accordingly. Therefore, in general, the design of multi-frequency operation and heterodyne detection makes NC-HEsFM an ideal tool to measure the electrostrain *via* a non-contact scheme with the capability of automatic resonance tracking.

**Experimental set-up.** The model NC-HEsFM system is constructed on a commercial Scanning Probe Microscope (SPM) system (SPA400, Seiko Instruments). In order to use QTF as the force sensor for non-contact scanning, both the original height feedback circuit and the probe holder of the SPM system have been modified. The complete set-up of this model system is schematically shown in Fig. 1*A*. A conductive AFM cantilever is glued at the end of the QTF prong using conductive silver paste (Fig. 1*B*), and then the QTF sensor is mounted on a piezo transducer with a small dip angle (Fig. 1*C*). The dip angle guarantees that the nano-tip on the cantilever end is located at the forefront of the sensor when engaging to the sample surface (Fig. 1*C*). As the additional mass induced by the AFM cantilever and the silver paste is very small, the resonance of the QTF can still keep a high *Q* factor after tip attached. Fig. 1*D* and *E* respectively show the typical free resonance curves of the 1$^{st}$ and 2$^{nd}$ flexural mode (anti-symmetric vibration) measured from the QTF with the AFM tip attached, in which a *Q* factor of 4000 ~ 6000 (8000 ~ 10000) in the 1$^{st}$ (2$^{nd}$) flexural mode is usually obtained. Although both the 1$^{st}$ and 2$^{nd}$ flexural mode can be used for topography measurement (90, 91), in the model NC-HEsFM system, only the 2$^{nd}$ flexural mode is employed to measure topography while the 1$^{st}$ flexural mode is used to detect the DFE signal because of its higher sensitivity. In the NC-HEsFM designed here, the QTF electrode with tip connected is set to be grounded while the drive signal for stimulating electrostrain is sent to the conductive substrate of the sample. During measurement, the QTF is mechanically excited to vibrate at its 2$^{nd}$ and a higher $n^{th}$ ($n \geq 3$) flexural mode (anti-symmetric vibration) simultaneously by the holder drive at frequency $f_2$ and $f_h$ respectively, while the sample electrostrain is stimulated by the sample drive at frequency $f_s$. To maximize the SNR of the DFE signal, the difference frequency $f_{diff}$ is set to near the resonance frequency of the 1$^{st}$ flexural mode. The deflection signal which contains all the QTF vibration information is obtained by amplifying the QTF current through a current amplifier located close to the QTF. A phase locked loop (PLL) established on a lock-in amplifier and a voltage-controlled oscillator (VCO) is employed to track the resonance frequency shift ($\Delta f_2$) of the 2$^{nd}$ flexural mode, and the $\Delta f_2$ is used as the height feedback signal for constant frequency shift scanning. Typically, under the ambient conditions, the tip is much easier to work with a positive frequency shift setpoint ($\Delta f_{sp}$) (78, 85, 90), while in the high vacuum environment, a negative $\Delta f_{sp}$ is usually used to achieve atomic resolution imaging (76, 80). As the effective $Q$ factor of the QTF changes dramatically when the tip comes close to the sample, an automatic gain control (AGC) is designed here to keep the vibration amplitude of the 2$^{nd}$ flexural mode at a constant setpoint $A_{sp}$. For the closed-loop feedback control, the classic proportion-integration (PI) strategy is adopted for both PLL and AGC. The DFE signal is demodulated by using a lock-in amplifier, in which the difference-frequency reference signal is produced by a home-made analog multiplier and a low-pass filter (LPF). Alternatively, the reference signal can be provided internally by synchronizing the clocks of signal source and lock-in amplifier (63). The demodulation results, i.e., the amplitude and phase of the DFE signal, are sent to the controller of SPM and then synchronously imaged with the topography (all the amplitude images here are shown in dimension of a.u.).

**Ferroelectric characterization.** To test the model NC-HEsFM system, several typical ferroelectric materials, including periodically poled lithium niobate (PPLN) single crystal, Pb(Zn$_{1/3}$Nb$_{2/3}$)O$_3$-9%PbTiO$_3$ (PZN-9%PT) single crystal, Pb(Zr,Ti)O$_3$ (PZT) film and BiFeO$_3$ (BFO) film are selected here as the standard sample since the electrostrain contrast (dominated by inverse piezoelectric effect) on those samples is deterministic and well defined. In the following experiments, the PtIr5 coated AFM probe (PPP-NCSTPt, Nanosensors) is used as the conductive tip by transferring the whole cantilever part to the QTF prong. During the NC-HEsFM measurement, a $\Delta f_{sp}$ within +20 ~ +50 Hz is set for the height feedback control and the $A_{sp}$ is set to be 2 ~ 10 mV. The AC drives for sample and the $n^{th}$ flexural mode are adjusted in the experiment according to the observed DFE signal strength, typically a drive amplitude of 2 ~ 8 V$_{pp}$ and 0.2 ~ 10 V$_{pp}$ is used for exciting sample and the $n^{th}$ flexural mode, respectively. All the experiments are performed under ambient conditions. Fig. 2*A*-*C* show the typical scanning results of the PPLN by NC-HEsFM. It is initially surprised that the simultaneously measured topography (Fig. 2*A*) shows many particle-shaped fine structures on the PPLN surface (see more data in Supplementary Materials, Fig. S2), which has been rarely revealed by conventional contact-mode PFM in the previous studies (62, 94-96). These fine structures are most probably the residuals that come from the microfabrication process of the PPLN sample (97). As NC-HEsFM is operated in a true non-contact mode, the surface damages or modifications have been eliminated thus the original surface status, including many details, can be well-preserved and then defined by the nano-tip. A detail comparison between conventional PFM and NC-HEsFM with respect to the surface modification will be discussed later. From the amplitude and phase images of the DFE signal, the domain walls between the two adjacent domains can be clearly observed in the amplitude image (Fig. 2*B*), and the periodical domains with alternative upward and downward polarization are distinctly revealed in the phase image (Fig. 2*C*). Meanwhile, Fig. 2*B* and 2*C* indicate an almost uniform amplitude distribution and a nearly 180° phase difference between the domains with opposite polarization, which agrees well with the characteristics of the proposed "ideal" ferroelectric characterizations (59, 98). Another ferroelectric sample, PZN-9%PT single crystal, is also studied by NC-HEsFM. The PZN-9%PT sample tested here has spontaneously polarized domains in which the polarization is along thickness direction, and the upward and downward domains are randomly distributed (99). Typical topography and simultaneously obtained electrostrain images of the PZN-9%PT sample are shown in Fig. 2*D*-*F*. Clear labyrinthine domain pattern and ~180° phase difference between the upward and downward domains can be observed, which are highly consistent with the reported results (62, 99-101), thereby once again confirming the validity of NC-HEsFM in ferroelectric domain characterization.

Similar to the conventional PFM, NC-HEsFM also allows to manipulate the ferroelectric domain by applying DC electric field. Fig. 2*G*-*I* show the structure of an artificially written domain on PZT film. The circular domain is created by maintaining the tip at the near surface (feedback on) while applying a -8 V DC voltage to sample (tip grounded) for ~10 s, then this artificial domain is *in-situ* imaged by NC-HEsFM. The clear domain contrast shown in Fig. 2*H* and 2*I* indicates that NC-HEsFM can unambiguously perform similar ferroelectric lithography as conventional PFM but with a non-contact manner. Another important observation is that the NC-HEsFM can theoretically break the sub-10 nanometers resolution limit of the conventional PFM, to demonstrate this, a high-resolution scanning (10 × 10 nm$^2$), has been conducted around the domain wall area of the artificial domain (indicated by the white box of Fig. 2*H*), and the results are displayed in Fig. 2*J-L*. From the DFE amplitude (Fig. 2*K*) and phase (Fig. 2*L*) images, it can be seen that the position of zero amplitude and 180° phase jump are defined within a ~2 nm range (indicated by the white dashed lines). Note that some thermal drift may exist in between the amplitude and phase images, as they are sequentially acquired by performing *in-situ* scanning twice (due to the limited signal channel of the SPA400 system). Theoretically, when the nano-tip is scanning over the narrow domain wall (estimated to be ~1 nm wide (102)), the amplitude and phase signal will correspondingly drop to zero and jump 180° respectively, implying that the zero amplitude and 180° phase jump can reflect the position of the domain wall. However, in conventional PFM, the piezoresponse signal usually has a considerable contribution from the local and distributed electrostatic force, causing the zero amplitude and 180° phase jump cannot reflect domain wall information accurately (103). In contrast, the electrostatic force, including both the local and distributed part, has been significantly minimized in NC-HEsFM (will be discussed later), thereby the observed ~2 nm range of zero amplitude and 180° phase jump (Fig. 2*K,L*) indicate that the domain wall position can be determined with a lateral resolution of ~2 nm. To the best of our knowledge, such high-resolution domain wall images have hardly been reported in previous by using the conventional PFM in the ambient environment. However, the results of Fig. 2*J-L* does not reflect the resolution limit of NC-HEsFM as the current model system has quite strong thermal noise, thermal drift and feedback noise (can be observed from the topography image, Fig. 2*J*), and at the same time, the sample surface and tip as well as the environment conditions do not well support high-resolution imaging. According to the prediction of Kalinin et al. (12, 15), it is reasonable to expect that, under the conditions where atomic resolution imaging can be easily achieved by QTF-based NC-AFM (76, 80), NC-HEsFM should be able to similarly provide an atomically resolved characterization for ferroelectric domain due to its superior capabilities of non-contact detection and electrostatic force minimization.

In addition, the measurement of ferroelectric hysteresis loop (switching spectroscopy) by using NC-HEsFM is also demonstrated here. The switching spectroscopy in NC-HEsFM is similar with that in the conventional PFM (53, 61, 104). A continuous or pulsed triangular wave-like DC bias sequence is superimposed on the AC drive and then applied to the sample to induce the polarization switching, while at the same time, measuring the DFE signal as a function of DC bias. Here, a continuous triangular DC probing wave with a step duration of 2 ms is employed to measure the hysteresis loop of the PZT film, which is similar with the macroscale polarization-electric field hysteresis loop measurement (105). Fig. 2*M* displays the attained hysteresis loops of PZT film. The NC-HEsFM amplitude loop (top) and phase loop (middle) show the expected butterfly shape and dual 180° phase jumps respectively, which are the standard characteristics of ferroelectricity (19, 50, 105). Meanwhile, the DFE in-phase (bottom), calculated by [amplitude × cos(phase)], manifests the typical ferroelectric hysteresis with applied DC bias (19, 50, 105). Intriguingly, the DFE phase loop shows a non-symmetric evolution trend where a ~180° phase difference can be observed between two polarization switching points, which is quite different with the symmetric phase loop obtained by using the conventional PFM (44, 50). This interesting non-symmetric phase evolution may indicate that the DFE phase signal can differentiate the two opposite polarization switching states (104, 106), while the mechanism is pending for further investigation. Note that, instead of using the pulsed DC method to minimize the electrostatic force effect in the conventional PFM (16, 53, 104, 107), the continuous DC method is employed here to acquire hysteresis loops. The obtained results, however, does not show observable features of the electrostatic force effect (50, 53, 108), this improvement in fact benefits from the NC-HEsFM's powerful electrostatic force suppression capability which will be discussed later.

Similar experiments, including both domain writing and switching spectroscopy, have been conducted on BFO film and the results are displayed in Supplementary Materials, Fig. S3. In brief, the results shown in Fig. 2 and Fig. S3 well demonstrate that NC-HEsFM can provide standard ferroelectric domain characterization, domain manipulation and hysteresis loop measurement.

**Multi-eigenmode operation.** In addition, NC-HEsFM can be operated at multiple eigenmodes. As the QTF has theoretically an infinite number of flexural modes, all the flexural modes with mode number $n \geq 3$ are supposed to be able to achieve the heterodyne detection of the electrostrain in NC-HEsFM if each eigenmode can be effectively excited. Therefore, in this work, six flexural modes (anti-symmetric vibration) with mode number from 3 to 8 are employed to measure the domain structure of PPLN, and the results are displayed in Fig. 3$A$ to 3$F$, respectively. Finite element simulation (see simulation details in Supplementary Materials, S6) is implemented to obtain the mode shape (shown in the first row of Fig. 3) and eigenfrequency of the each mode, and the experimental determined eigenfrequencies agree well with the simulation results (Supplementary Materials, Table S1). From Fig. 3, it is obvious that all of the six flexural modes can be used for electrostrain measurement, and the results, including DFE amplitude and phase images, attained using each flexural mode manifest an almost same contrast of the domains. Note that nonuniform distribution of DFE amplitude can be observed on several amplitude images (Fig. 3$A,E,F$), this is most likely caused by a background difference-frequency interference as it is found that there is a difference-frequency component generated in the current amplifier due to its non-linear effect (see details in Supplementary Materials, S2). By applying appropriate modifications to the original current amplifying circuit, such as adding low-pass filters, the difference-frequency component can be minimized. The operability of multiple eigenmodes implies that, when necessary, the excitation frequency can be adjusted *via* mode selection. For instance, when using NC-HEsFM to measure electrochemical Vegard strain, the $3^{rd}$ flexural mode may be used to obtain a stronger signal as a lower excitation frequency corresponds to a larger Vegard strain (65). Whereas if the sample under test is prone to be affected by electrochemical Vegard strain or dynamic electrochemical reactions, the $8^{th}$ or even higher flexural mode can be used since higher excitation frequency can help to minimize these influences (8, 61, 62).

**Non-destructive characterization.** It is generally believed that the conventional contact-mode PFM is a non-destructive method for nanoscale piezo/ferro-electric characterization (37, 109). In fact, however, the contact-mode scanning can be extremely prone to introduce modifications to both the sample surface and tip apex due to the relatively large and continuous tip-sample repulsive interaction, indicating that it is very difficult to maintain a constant tip and sample status during the entire PFM scanning process. Therefore, the contact-mode operation of the conventional PFM usually causes a lot of challenges to the reproducibility of the final topography and PFM images. Herein, the as-purchased fresh PPLN sample has been used to demonstrate the destructive yet irreversible scanning by the conventional PFM. PPLN is a standard sample which has been extensively used to calibrate the PFM signal, and many PFM images of PPLN including the simultaneously obtained topographies have been reported (62, 94-96). Similar results can also be obtained in our experiments when performing the conventional PFM measurements (see Supplementary Materials, S3). However, when PPLN is measured by the newly developed NC-HEsFM, it is surprised at the beginning to see so many fine structures on the surface (Fig. 2*A*), which has seldom been revealed previously by PFM on commercial PPLN samples. To investigate the possible reasons, the tapping-mode AFM and PFM are used together to measure the topography. Firstly, tapping-mode AFM scanning is executed on a fresh area with a new tip to obtain the original surface topography, then *in-situ* PFM scanning is performed on a zoomed-in area (indicated by the box of Fig. 4*A*) for a few times to get the contact-mode topography images, and finally, tapping-mode AFM is used again to check the final surface status. Fig. 4*A*-*H* show the surface topography images measured by tapping-mode AFM and conventional PFM on PPLN using an AFM probe with stiffness of ~2 N/m (240AC-PP, OPUS).

Obviously, the initial topography images (Fig. 4A,B) does clearly show the fine structures on PPLN surface. However, the topography images obtained by PFM scanning (Fig. 4C-F) are remarkably different with the tapping-mode topography shown in Fig. 4B, and those fine features, even including the large protrusions, have gradually been "cleaned" during the contact-mode scanning. This can be consistent with the frequently observed phenomenon that the PFM images usually manifest better and better when continuously scanning on a same area (see Supplementary Materials, Fig. S5). Fig. 4G and 4H unambiguously demonstrate what happening during the contact-mode PFM scanning. Comparing Fig. 4G,H with Fig. 4A,B, it is evident that both the fine and large protrusion structures have been scratched off by the tip, well indicating that the contact-mode PFM scanning can be highly destructive. Therefore, the original fine features of the PPLN surface can hardly be observed in general PFM measurements (62, 94-96). In contrast, situations are dramatically different in NC-HEsFM measurements. Similar experiments have been conducted on PPLN using NC-HEsFM, and the results are presented in Fig. 4I-P. Fig. 4I,J show the initial scanning results, in which the original fine structures have been well defined in the topography images and no noticeable changes can be observed between the first (Fig. 4I) and second (Fig. 4J) scanning. Continuous zoomed-in scanning has been conducted *in-situ* to check if any modifications emerge and the results are shown in Fig. 4K-O. Fig. 4K,L and Fig. 4M,N show the obtained topography images corresponding to the scanning areas indicated in the Fig. 4J and Fig. 4L respectively. Clearly, no any observable surface modification can be found during the continuous non-contact scanning. To verify that the fine structures on PPLN surface can be easily damaged, a very weak indentation has been applied by the tip at the point shown in Fig. 4M. Fig. 4O shows the *in-situ* topography image (the box area in Fig. 4M) obtained after the indentation experiment. A triangular pyramid indentation can be clearly observed in Fig. 4O, implying that this surface layer indeed has a weak mechanical strength and that's why it can be easily damaged by the contact-mode scanning. After many times zoomed-in scanning with the NC-HEsFM, a final scanning has been performed within the initial area to check the surface status and the result is displayed in Fig. 4P. Comparing Fig. 4I,J with Fig. 4P, nearly no zoomed-in scanning traces and surface modifications can be observed, which unambiguously indicates that the NC-HEsFM has achieved a real non-destructive measurement. Several areas with large protrusion structures have also been selected to test the NC-HEsFM and the results are presented in Supplementary Materials, Fig. S6. Meanwhile, a much stricter and longer experiment which includes 16 times *in-situ* scanning (last for ~9 hours) has been conducted to test the stability and reproducibility of NC-HEsFM images (see Supplementary Materials, S4). All the results clearly demonstrate that the NC-HEsFM can stably provide a non-contact, non-destructive and highly reproducible simultaneous topography and electrostrain measurement. Since both the tip and sample surface can be largely protected in NC-HEsFM, the initial tip and sample status, especially for atomically sharp tip and atomically flat surface, now can be well kept during the scanning, which provides an ideal condition for achieving ultrahigh-resolution imaging (12, 15). However, such ultrahigh-resolution imaging usually requires strict measurement conditions and environments, typically the high vacuum or even low temperature, and moving the NC-HEsFM system to a vacuum chamber needs many modifications thus this part of study will be revealed in the future.

**Electrostatic force effect.** Electrostatic force is well-known a great yet long-standing challenge for conventional PFM. As discussed earlier, huge efforts have been made to address this intractable problem, however the proposed solutions are still subject to many limitations due to the technique principle of the PFM. The invention of HM-PFM breaks the conventional technique frame and provides a brand-new pathway to minimize the electrostatic force by using heterodyne detection and high-frequency excitation. The mechanism of minimizing the electrostatic force effect in HM-PFM has been systematically discussed in our previous study (62), and in brief, it is based on two aspects, one is the significantly increased effective stiffness of the cantilever at high eigenmodes, and another is the heterodyne detection which breaks the direct coupling between the electrostrain and electrostatic force signals. In NC-HEsFM, the 1$^{st}$ flexural mode of the QTF has an stiffness as high as ~1800 N/m (76, 89), which is already far larger than that of any conventional cantilever, let alone the $n^{th}$ ($n \geq 3$) flexural mode where the effective stiffness increases sharply with the mode number $n$ (62). Therefore, during NC-HEsFM measurement, the QTF will certainly has a large enough stiffness to minimize both the local and distributed electrostatic force effects. Meanwhile, the heterodyne detection scheme has also been used in the NC-HEsFM, thus with the basis of HM-PFM, NC-HEsFM should achieve a similar or even much better of electrostatic force minimization. To confirm this capability, the widely used electrostatic force examining method, i.e., the DC spectroscopy (47-49), is conducted on PPLN by conventional PFM and NC-HEsFM. Fig. 5*A* and 5*B* show the PFM amplitude and phase as a function of DC bias, respectively. In Fig. 5*A*,*B*, the typical electrostatic force-induced V-shaped amplitude curve and ~180° phase change can be observed clearly, implying that the electrostatic force is significantly affecting the PFM signal (47-49). In contrast, completely different variation trends of the DFE amplitude and phase can be observed in the measurement of NC-HEsFM (Fig. 5*C*,*D*). It is evident that, even though a large DC bias scanning range (±30 V) is used, all of the DFE amplitude and phase signals measured by six different flexural modes can keep a constant magnitude with changing DC bias, which unambiguously indicates that the electrostatic force effect, including both the local and distributed part, has almost been completely eliminated in the NC-HEsFM. With this superior capability of minimizing electrostatic force effect, the domain wall position can thereby be determined accurately from the DFE amplitude and phase images (Fig. 2K,L) and at the same time, standard ferroelectric hysteresis loop (Fig. 2M) can be obtained even though continuous DC method is used in the switching spectroscopy measurement.

**Mechanism of effective electrostrain measurement.** Above results clearly show that NC-HEsFM can effectively detect the tiny electrostrain with a true non-contact manner. Therefore an obvious question is why NC-HEsFM can realize an effective electrostrain measurement while the intermittent-contact mode PFM (72) cannot. As discussed in the previous section, the key factor is the tip-sample distance because it dominates the electric field distribution and tip-sample interaction force. To show how tip-sample distance ($z$) affects the electric field distribution, a finite element simulation has been employed here to calculate the electric field distributions under different values of $z$, and the results are shown in Fig. 6A,B (see simulation details in Supplementary Materials, S6). Fig. 6A and 6B shows the cross-sectional and surface electric field distributions with $z = 0.1$, 2.1 and 4.1 nm, respectively. Meanwhile, the line profiles (indicated in Fig. 6B) of the surface electric field strength under a series of $z$ are plotted in Fig. 6C. From Fig. 6A-C, it is obvious that the effective electric field applied to the sample dramatically decays with increasing tip-sample separation even though the tip is only lifted by a few nanometers. Therefore, when using conventional cantilever-based tip to measure electrostrain with non-contact or intermittent-contact mode, the large tip vibration (typically with amplitude of a few to tens of nanometers) can hardly stimulate a relatively stable and large enough surface strain. In addition to the electric field, the tip-sample interaction is also of great importance for effectively detecting the electrostrain since the tiny surface strain is measured indirectly based on a strong enough tip-sample interaction. Here, for simplicity, the well-studied Si-Si (Si tip on Si surface) interaction model (73) has been employed to demonstrate the relationship between tip-sample interaction force and $z$. Fig. 6D shows the calculated 1$^{st}$-order tip-sample force gradient as a function of $z$ under the assumption of ideal tip and surface (see calculation details in Supplementary Materials, S5). Since the tip-sample force gradient determines the coupling between sample surface and tip when measuring electrostrain (15, 110), such a sharp attenuation trend of the force gradient shown in Fig. 6D clearly indicates that even lifting the tip by ~1 nm, the force gradient thereby the sensitivity for surface displacement can decrease dramatically. Therefore, using cantilever-based tip to realize non-contact or intermittent-contact detection of the electrostrain is very difficult. From Fig. 6*A-D*, it can be speculated that if the dynamic tip-sample separation is well controlled within the range of angstrom, there should be a large possibility to achieve effective electrostrain measurement, and in fact, this is exactly the main mechanism used in the NC-HEsFM measurements. To confirm this, the practical tip vibration amplitude of NC-HEsFM has been estimated using the thermal noise analysis method (89, 111). Fig. 6*E* and *6F* show the typical power spectral density (PSD) spectrums of the output voltage noise (from current amplifier) acquired around the eigenfrequency of the 1$^{st}$ and the 2$^{nd}$ flexural mode, respectively. By defining a voltage deflection sensitivity $\alpha$, the voltage noise PSD of an eigenmode can be expressed as (see details in Supplementary Materials, S7):

$$S_{VV}(f) = \alpha^2 \frac{4k_B T f_0^3 / (\pi k_0 Q)}{\left(f_0^2 - f^2\right)^2 + \left(f_0 f / Q\right)^2} \quad (4)$$

where $k_B$, $T$, $f_0$ and $k_0$ are Boltzmann constant, temperature, resonance frequency and mode stiffness, respectively. By fitting the PSD spectrums of the two modes using Equation (4) and the constants of $T = 298.15$ K, $k_0 = 1800$ N/m (1$^{st}$ flexural mode) and 70740 N/m (2$^{nd}$ flexural mode), the voltage deflection sensitivity of the 1$^{st}$ and the 2$^{nd}$ flexural mode can be obtained, which are 21 and 62 µV/pm (for two-arm anti-symmetric vibration), respectively. With the deflection sensitivity, the tip vibration amplitudes of the NC-HEsFM can be estimated to be 30 ~ 160 pm (2$^{nd}$ flexural mode) and 20 ~ 40 pm (1$^{st}$ flexural mode) in this study. For the $n^{th}$ high flexural mode, the vibration amplitude cannot be estimated by this thermal noise method as its PSD spectrum manifests below the noise floor. However, the amplitude of the $n^{th}$ flexural mode should be close to or even smaller than that of the 2$^{nd}$ flexural mode since during the normal NC-HEsFM measurement, no change can be observed in both PLL and AGC feedbacks when the high mode drive is turned on/off. Using the operation condition of $\Delta f_{sp}$ (+20 ~ +50 Hz), the average force gradient during a vibration period can be estimated to be 14 ~ 36 N/m (under small amplitude approximation (76)). According to the estimated amplitudes and average force gradient, it can be concluded that, during the whole vibration period, the tip-sample distance in NC-HEsFM can be well controlled within angstrom or even sub-angstrom range, which can fully guarantee the strength of both sample electric field and tip-sample interaction thus effective electrostrain measurement can be easily reached in NC-HEsFM.

## Conclusion and Outlook

In this study, an advanced scanning probe technique called Non-Contact Heterodyne Electrostrain Force Microscopy (NC-HEsFM) has been introduced, for the first time, to measure the nanoscale electrostrain with true non-contact manner. With using quartz tuning fork (QTF) force sensor and heterodyne detection scheme, NC-HEsFM firstly achieves the long-awaited goal of non-contact ferroelectric characterization at nanoscale. It has been unambiguously demonstrated that NC-HEsFM can perform ideal high-resolution ferroelectric domain mapping, standard ferroelectric hysteresis loop measurement and controllable domain manipulation. At the same time, NC-HEsFM is able to operate on multiple high eigenmodes thus the frequency of sample excitation can be adjusted to adapt the materials and measurement demands. By comparing with the measurements using conventional PFM, NC-HEsFM shows an unprecedented capability in achieving real non-destructive ferroelectric characterization with significantly minimized local and distributed electrostatic force effects. The qualitative analysis based on theoretical model and finite element simulations indicates that the effective electrostrain measurement achieved in NC-HEsFM stems from the highly controllable angstrom or even sub-angstrom tip-sample separation offered by the QTF sensor, because such a small tip-sample separation fully guarantees the strength of both sample electric field and tip-sample interaction required for performing nanoscale electrostrain measurement. In brief, by introducing QTF force sensor, heterodyne detection and multi-frequency operation scheme, NC-HEsFM has simultaneously overcome the most fundamental and long-standing issues such as contact-mode operation, resonance tracking and electrostatic force faced by conventional PFM, providing a significantly improved, non-contact yet non-destructive, automatically resonance tracked and electrostatic force minimized way for nanoscale ferroelectric or piezoelectric studies.

In addition, the small volume and piezoelectric self-detection (i.e., non-optical detection) of the QTF force sensor make NC-HEsFM an ideal candidate for measuring nanoscale electrostrain under extreme environments like ultra-high vacuum and low temperature, which may largely support the studies towards investigating those most intrinsic ferro/piezo-electric phenomena, such as the ferroelectric polarization dynamics (16). Given the ultra-high spatial resolution achieved in the QTF-based NC-AFM, it is believed that, under appropriate conditions such as high vacuum, NC-HEsFM may similarly provide the ultra-high resolution imaging for electrostrain measurement. Meanwhile, the ultra-stiff QTF prong enables NC-HEsFM to perform *in situ* STM or tunneling spectroscopy measurements, which can have great significance for studying the ferro/piezo-electricity of ultrathin films (2, 88, 112, 113) and conductive ferroelectric domain walls (114, 115). Furthermore, NC-HEsFM can be simply modified to measure the high-order electrostrains (see Supplementary Materials, S8) (63), such as the $2^{nd}$-order electrostrictive strain (67), in which the contact-mode and the $2^{nd}$-harmonic electrostatic force issues can be avoided comparing to that of the currently-used PFM-based method (68).

Finally, the success of the NC-HEsFM in measuring nanoscale electrostrain can be easily transplanted to multiple other surface strain-based scanning probe microscopies (63), such as the ultrasonic strain-based Atomic Force Acoustic Microscopy (116), the photothermal strain-based Infrared-AFM or photothermal induced resonance technique (117, 118), the magnetomechanical strain-based Piezomagnetic Force Microscopy (119) and the Joule thermal strain-based Scanning Joule Expansion Microscopy (120). Due to the similarities in the technical principles and architectures, all of these scanning probe techniques are faced with similar fundamental problems as those in the conventional PFM, including the contact-mode operation and the $1^{st}$-harmonic interference (such as electrostatic force and photo-induced force). With the introduction of NC-HEsFM, these fundamental issues can be well solved by using similar technical design, thereby in general, bringing significantly improved characterization capabilities for studying the nanoscale surface strains, as well as the corresponding physical properties, under the excitation of electric, acoustic, optical, magnetic and temperature fields (63).

## Materials and Methods

**Experimental set-up.** The model NC-HEsFM system was established on a commercial SPM (SPA400, Seiko Instruments), and hardware modifications were implemented to enable external height feedback control. The original tip holder was redesigned for the application of QTF sensor. The AC excitation signals were generated by an arbitrary waveform generator (Keysight 33522B, Keysight Technologies). An embedded field programmable gate array (FPGA) controller (NI cRIO-9064, National Instruments) equipped with one digital acquisition card (NI 9775, National Instruments) and one analog output card (NI 9263, National Instruments) was utilized to complete the signal acquisition, PLL and AGC feedback control. A current amplifier (DLPCA-200, FEMTO) with the gain of $1\times10^9$ V/A was employed to amplify the QTF current signal. The demodulation of the AC signals was finished by using a multi-channel lock-in amplifier (MFLI, Zurich Instruments). The home-made functional circuits, including analog multiplier, low pass filter and summing amplifier, were used to complete AC signal summation and AGC amplitude control as well as generate the difference-frequency reference signal. DC bias was provided by a digital source meter (Keithley 2450, Tektronix), and the superposition of DC bias to AC drive was finished by a bias-tee. All the programmable instruments were controlled by the self-developed NC-HEsFM control program based on LabVIEW$^{TM}$ 2017 and LabVIEW$^{TM}$ FPGA 2017 (National Instruments).

**SPM characterization.** A QTF (Type E158, Micro Crystal) with fundamental resonance frequency of 32.768 kHz was used in this study as the force sensor (63). A PtIr5 coated conductive AFM probe (PPP-NCSTPt, Nanosensors) was employed here as the scanning tip, of which the cantilever was glued at the end of the QTF prong using conductive silver paste. When the AFM tip is disabled, it can be simply removed from the QTF with the aid of organic solvent and a new tip can be mounted again (63). During the scanning, the frequency shift setpoint for height feedback control was typically set to +20 ~ +50 Hz and the scanning line speed was 0.25 ~ 0.75 lines/s. Due to the limited signal acquisition channel of the SPA400 system, the amplitude and phase images of NC-HEsFM were obtained by executing *in-situ* scanning twice. The drive amplitudes for the higher flexural mode and sample were typically set to 0.05 ~ 10 $V_{pp}$ and 2 ~ 8 $V_{pp}$ respectively. For the conventional PFM measurement, a Pt coated conductive AFM probe (240AC-PP, OPUS) with a force constant of ~2 N/m and free resonance frequency of ~70 kHz was used. All measurements were conducted in ambient environment with tip grounded. Although this study is conducted in ambient environment due to the limitation of SPA400 system, it is highly recommended to perform NC-HEsFM measurements, especially the ferroelectric domain manipulation and hysteresis loop measurement, in noble gas atmosphere or vacuum because it is found that these measurements are prone to induce surface electrochemical reactions under ambient conditions (121).

**Sample preparation.** A commercially available PPLN standard sample (AR-PPLN test sample, Asylum Research, Oxford Instruments) was used, which consists of a 3 mm × 3 mm transparent die with a thickness of 0.5 mm. The PZN-9%PT single crystal (supplied by Microfine Materials Technology Pte. Ltd., Singapore) was cut into small piece with rectangular shape and the respective orientations are $[100]^L/[010]^W/[001]^T$ and the surface of the samples was polished with SiC papers and alumina powder using water-cooled polisher. After the polishing processes, the dimension of the samples is approximately 4 mm (width) × 4 mm (length) × 0.5 mm (thickness). The PZT 20/80 film with thickness of ~300 nm was grown in SrRuO$_3$-buffered SrTiO$_3$ (STO) substrate with (001) orientation using pulsed laser deposition (PLD) (KrF excimer laser, λ = 248 nm). The SrRuO$_3$ layer (~50 nm) was firstly deposited on the SrTiO$_3$ substrate at a temperature of 680°C and an oxygen pressure of 15 Pa. Then the PZT layer was grown on top of the SrRuO$_3$ layer at a temperature of 600°C and the same 15 Pa oxygen pressure. After growth, the film was cooled to room temperature at 10°C/min in an oxygen atmosphere of 1 atm. For the preparation of BFO(100nm)/LSMO(20nm)/STO film, a commercial 10% Bi excess BFO target was used to compensate the Bi volatilization during the PLD growth process. The commercial La$_{0.33}$Sr$_{0.67}$MnO$_3$ target was used to grow the LSMO layer. The BFO/LSMO heterostructure was grown on STO substrate with (001) orientation using PLD at 700 °C under an oxygen partial pressure of 100 mTorr and cooled in a 1 atm oxygen atmosphere. The laser fluence and repetition rate were 1.3 J/cm$^2$ and 8 Hz, respectively.

**Finite element simulation.** The modal analysis of the QTF was finished by using the Structural Mechanics Module of the commercial software package COMSOL Multiphysics® 5.3 (COMSOL Inc.), while the tip electric field simulations were conducted using its AC/DC Module. For the purpose of simplification, a 2D axisymmetric model was built for the simulation of the electric field distribution. More simulation details can be found in Supplementary Materials, S6.

## References

1. Scott JF (2007) Applications of Modern Ferroelectrics. *Science* 315(5814):954-959.

2. Dawber M, Rabe KM, & Scott JF (2005) Physics of thin-film ferroelectric oxides. *Rev. Mod. Phys.* 77(4):1083-1130.

3. Gruverman A & Kholkin A (2006) Nanoscale ferroelectrics: processing, characterization and future trends. *Rep. Prog. Phys.* 69(8):2443-2474.

4. Grinberg I, *et al.* (2013) Perovskite oxides for visible-light-absorbing ferroelectric and photovoltaic materials. *Nature* 503:509.

5. Martin LW & Rappe AM (2016) Thin-film ferroelectric materials and their applications. *Nat. Rev. Mater.* 2(2):16087.

6. Güthner P & Dransfeld K (1992) Local poling of ferroelectric polymers by scanning force microscopy. *Appl. Phys. Lett.* 61(9):1137-1139.

7. Gruverman A, Alexe M, & Meier D (2019) Piezoresponse force microscopy and nanoferroic phenomena. *Nat. Commun.* 10(1):1661.

8. Collins L, Liu Y, Ovchinnikova OS, & Proksch R (2019) Quantitative Electromechanical Atomic Force Microscopy. *ACS Nano* 13(7):8055-8066.

9. Kalinin SV, *et al.* (2007) Intrinsic single-domain switching in ferroelectric materials on a nearly ideal surface. *Proc. Natl. Acad. Sci. U.S.A* 104(51):20204.

10. Steffes JJ, Ristau RA, Ramesh R, & Huey BD (2019) Thickness scaling of ferroelectricity in $BiFeO_3$ by tomographic atomic force microscopy. *Proc. Natl. Acad. Sci. U.S.A* 116(7):2413.

11. Jesse S, *et al.* (2008) Direct imaging of the spatial and energy distribution of nucleation centres in ferroelectric materials. *Nat. Mater.* 7(3):209-215.

12. Kalinin SV, *et al.* (2007) Nanoscale Electromechanics of Ferroelectric and Biological Systems: A New Dimension in Scanning Probe Microscopy. *Ann. Rev. Mater. Res.* 37(1):189-238.

13. Gruverman A & Kalinin SV (2006) Piezoresponse force microscopy and recent advances in nanoscale studies of ferroelectrics. *J. Mater. Sci.* 41(1):107-116.

14. Kwon O, Seol D, Qiao H, & Kim Y (2020) Recent Progress in the Nanoscale Evaluation of Piezoelectric and Ferroelectric Properties via Scanning Probe Microscopy. *Adv. Sci.* 7(18):1901391.

15. Kalinin SV, Rar A, & Jesse S (2006) A decade of piezoresponse force microscopy: progress, challenges, and opportunities. *IEEE Trans. Ultrason., Ferroelectr., Freq. Control.* 53(12):2226-2252.

16. Kalinin SV, Morozovska AN, Chen LQ, & Rodriguez BJ (2010) Local polarization dynamics in ferroelectric materials. *Rep. Prog. Phys.* 73(5):056502.

17. Kalinin SV, Kim Y, Fong DD, & Morozovska AN (2018) Surface-screening mechanisms in ferroelectric thin films and their effect on polarization dynamics and domain structures. *Rep. Prog. Phys.* 81(3):036502.

18. Chen B, *et al.* (2018) Large electrostrictive response in lead halide perovskites. *Nat. Mater.* 17(11):1020-1026.

19. Vasudevan RK, Balke N, Maksymovych P, Jesse S, & Kalinin SV (2017) Ferroelectric or non-ferroelectric: Why so many materials exhibit "ferroelectricity" on the nanoscale. *Appl. Phys. Rev.* 4(2):021302.

20. Seol D, Kim B, & Kim Y (2017) Non-piezoelectric effects in piezoresponse force microscopy. *Curr. Appl. Phys.* 17(5):661-674.

21. Zhou YS, *et al.* (2013) In Situ Quantitative Study of Nanoscale Triboelectrification and Patterning. *Nano Lett.* 13(6):2771-2776.

22. Balke N, *et al.* (2014) Exploring Local Electrostatic Effects with Scanning Probe Microscopy: Implications for Piezoresponse Force Microscopy and Triboelectricity. *ACS Nano* 8(10):10229-10236.

23. Kalinin A, Atepalikhin V, Pakhomov O, Kholkin AL, & Tselev A (2018) An atomic force microscopy mode for nondestructive electromechanical studies and its application to diphenylalanine peptide nanotubes. *Ultramicroscopy* 185:49-54.

24. Calahorra Y, Smith M, Datta A, Benisty H, & Kar-Narayan S (2017) Mapping piezoelectric response in nanomaterials using a dedicated non-destructive scanning probe technique. *Nanoscale* 9(48):19290-19297.

25. Ievlev AV, *et al.* (2018) Chemical Phenomena of Atomic Force Microscopy Scanning. *Anal. Chem.* 90(5):3475-3481.

26. Saadi MASR, Uluutku B, Parvini CH, & Solares SD (2020) Soft sample deformation, damage and induced electromechanical property changes in contact- and tapping-mode atomic force microscopy. *Surf. Topogr.: Metrol. Prop.* 8(4):045004.

27. Hurley DC, Kopycinska-Müller M, Kos AB, & Geiss RH (2005) Nanoscale elastic-property measurements and mapping using atomic force acoustic microscopy methods. *Meas. Sci. Technol.* 16(11):2167-2172.

28. Lantz MA, O'Shea SJ, & Welland ME (1998) Characterization of tips for conducting atomic force microscopy in ultrahigh vacuum. *Rev. Sci. Instrum.* 69(4):1757-1764.

29. Bhushan B & Kwak KJ (2008) The role of lubricants, scanning velocity and operating environment in adhesion, friction and wear of Pt–Ir coated probes for atomic force microscope probe-based ferroelectric recording technology. *J. Phys.: Condens. Matter* 20(32):325240.

30. Jesse S, *et al.* (2010) Resolution theory, and static and frequency-dependent cross-talk in piezoresponse force microscopy. *Nanotechnology* 21(40):405703.

31. Cui A, *et al.* (2019) Probing electromechanical behaviors by datacube piezoresponse force microscopy in ambient and aqueous environments. *Nanotechnology* 30(23):235701.

32. Lu H, *et al.* (2012) Mechanical Writing of Ferroelectric Polarization. *Science* 336(6077):59-61.

33. Kalinin SV, Jesse S, Liu W, & Balandin AA (2006) Evidence for possible flexoelectricity in tobacco mosaic viruses used as nanotemplates. *Appl. Phys. Lett.* 88(15):153902.

34. Wang L, *et al.* (2020) Flexoelectronics of centrosymmetric semiconductors. *Nat. Nanotech.* 15(8):661-667.

35. Kholkin AL, *et al.* (2003) Stress-induced suppression of piezoelectric properties in PbTiO3:La thin films via scanning force microscopy. *Appl. Phys. Lett.* 82(13):2127-2129.

36. Abplanalp M, Fousek J, & Günter P (2001) Higher Order Ferroic Switching Induced by Scanning Force Microscopy. *Phys. Rev. Lett.* 86(25):5799-5802.

37. Soergel E (2011) Piezoresponse force microscopy (PFM). *J. Phys. D: Appl. Phys.* 44(46):464003.

38. Jesse S, Mirman B, & Kalinin SV (2006) Resonance enhancement in piezoresponse force microscopy: Mapping electromechanical activity, contact stiffness, and Q factor. *Appl. Phys. Lett.* 89(2):022906.

39. Harnagea C, Alexe M, Hesse D, & Pignolet A (2003) Contact resonances in voltage-modulated force microscopy. *Appl. Phys. Lett.* 83(2):338-340.

40. Okino H, Sakamoto J, & Yamamoto T (2003) Contact-Resonance Piezoresponse Force Microscope and Its Application to Domain Observation of Pb($Mg_{1/3}Nb_{2/3}$)$O_3$–$PbTiO_3$ Single Crystals. *Jpn. J. Appl. Phys.* 42(Part 1, No. 9B):6209-6213.

41. Rodriguez BJ, Callahan C, Kalinin SV, & Proksch R (2007) Dual-frequency resonance-tracking atomic force microscopy. *Nanotechnology* 18(47):475504.

42. Jesse S, Kalinin SV, Proksch R, Baddorf AP, & Rodriguez BJ (2007) The band excitation method in scanning probe microscopy for rapid mapping of energy dissipation on the nanoscale. *Nanotechnology* 18(43):435503.

43. Jesse S & Kalinin SV (2011) Band excitation in scanning probe microscopy: sines of change. *J. Phys. D: Appl. Phys.* 44(46):464006.

44. Qiao H, Kwon O, & Kim Y (2020) Electrostatic effect on off-field ferroelectric hysteresis loop in piezoresponse force microscopy. *Appl. Phys. Lett.* 116(17):172901.

45. Seol D, Kang S, Sun C, & Kim Y (2019) Significance of electrostatic interactions due to surface potential in piezoresponse force microscopy. *Ultramicroscopy* 207:112839.

46. Eliseev EA, *et al.* (2014) Electrostrictive and electrostatic responses in contact mode voltage modulated scanning probe microscopies. *Appl. Phys. Lett.* 104(23):232901.

47. Kim S, Seol D, Lu X, Alexe M, & Kim Y (2017) Electrostatic-free piezoresponse force microscopy. *Sci. Rep.* 7:41657.

48. MacDonald GA, DelRio FW, & Killgore JP (2018) Higher-eigenmode piezoresponse force microscopy: a path towards increased sensitivity and the elimination of electrostatic artifacts. *Nano Futures* 2(1):015005.

49. Gomez A, Puig T, & Obradors X (2018) Diminish electrostatic in piezoresponse force microscopy through longer or ultra-stiff tips. *Appl. Surf. Sci.* 439:577-582.

50. Balke N, *et al.* (2015) Differentiating ferroelectric and nonferroelectric electromechanical effects with scanning probe microscopy. *ACS Nano* 9(6):6484-6492.

51. Balke N, *et al.* (2016) Quantification of surface displacements and electromechanical phenomena via dynamic atomic force microscopy. *Nanotechnology* 27(42):425707.

52. Balke N, *et al.* (2017) Quantification of in-contact probe-sample electrostatic forces with dynamic atomic force microscopy. *Nanotechnology* 28(6):065704.

53. Hong S, *et al.* (2001) Principle of ferroelectric domain imaging using atomic force microscope. *J. Appl. Phys.* 89(2):1377-1386.

54. Kim B, Seol D, Lee S, Lee HN, & Kim Y (2016) Ferroelectric-like hysteresis loop originated from non-ferroelectric effects. *Appl. Phys. Lett.* 109(10):102901.

55. Seal K, Jesse S, Rodriguez BJ, Baddorf AP, & Kalinin SV (2007) High frequency piezoresponse force microscopy in the 1-10 MHz regime. *Appl. Phys. Lett.* 91(23):232904.

56. Rodriguez BJ, Jesse S, Baddorf AP, & Kalinin SV (2006) High Resolution Electromechanical Imaging of Ferroelectric Materials in a Liquid Environment by Piezoresponse Force Microscopy. *Phys. Rev. Lett.* 96(23):237602.

57. Labuda A & Proksch R (2015) Quantitative measurements of electromechanical response with a combined optical beam and interferometric atomic force microscope. *Appl. Phys. Lett.* 106(25):253103.

58. Denning D, Guyonnet J, & Rodriguez BJ (2016) Applications of piezoresponse force microscopy in materials research: from inorganic ferroelectrics to biopiezoelectrics and beyond. *Int. Mater. Rev.* 61(1):46-70.

59. Proksch R (2015) In-situ piezoresponse force microscopy cantilever mode shape profiling. *J. Appl. Phys.* 118(7):072011.

60. Kim Y, *et al.* (2009) Origin of surface potential change during ferroelectric switching in epitaxial PbTiO$_3$ thin films studied by scanning force microscopy. *Appl. Phys. Lett.* 94(3):032907.

61. Zeng K & Zeng Q (2020) Singapore Non-Provisional Application No. 10202003740V.

62. Zeng Q, *et al.* (2021) Nanoscale Ferroelectric Characterization with Heterodyne Megasonic Piezoresponse Force Microscopy. *Adv. Sci.* DOI: 10.1002/advs.202003993.

63. Zeng K & Zeng Q (2021) Singapore Non-Provisional Application No. 10202100483P.

64. Gruverman A, Auciello O, & Tokumoto H (1996) Scanning force microscopy for the study of domain structure in ferroelectric thin films. *Journal of Vacuum Science & Technology B* 14(2):602-605.

65. Balke N, *et al.* (2010) Nanoscale mapping of ion diffusion in a lithium-ion battery cathode. *Nat. Nanotech.* 5(10):749-754.

66. Damjanovic D (1998) Ferroelectric, dielectric and piezoelectric properties of ferroelectric thin films and ceramics. *Rep. Prog. Phys.* 61(9):1267-1324.

67. Li F, Jin L, Xu Z, & Zhang S (2014) Electrostrictive effect in ferroelectrics: An alternative approach to improve piezoelectricity. *Appl. Phys. Rev.* 1(1):011103.

68. Yu J, *et al.* (2018) Quadratic electromechanical strain in silicon investigated by scanning probe microscopy. *J. Appl. Phys.* 123(15):155104.

69. Abdollahi A, Domingo N, Arias I, & Catalan G (2019) Converse flexoelectricity yields large piezoresponse force microscopy signals in non-piezoelectric materials. *Nat. Commun.* 10(1):1266.

70. Miao H, Tan C, Zhou X, Wei X, & Li F (2014) More ferroelectrics discovered by switching spectroscopy piezoresponse force microscopy? *EPL (Europhysics Letters)* 108(2):27010.

71. Kim Y, *et al.* (2011) Nonlinear Phenomena in Multiferroic Nanocapacitors: Joule Heating and Electromechanical Effects. *ACS Nano* 5(11):9104-9112.

72. Rodriguez BJ, Jesse S, Habelitz S, Proksch R, & Kalinin SV (2009) Intermittent contact mode piezoresponse force microscopy in a liquid environment. *Nanotechnology* 20(19):195701.

73. Giessibl FJ (1997) Forces and frequency shifts in atomic-resolution dynamic-force microscopy. *Phys. Rev. B* 56(24):16010-16015.

74. Giessibl FJ, Bielefeldt H, Hembacher S, & Mannhart J (1999) Calculation of the optimal imaging parameters for frequency modulation atomic force microscopy. *Appl. Surf. Sci.* 140(3):352-357.

75. Giessibl FJ (1995) Atomic resolution of the silicon (111)-(7x7) surface by atomic force microscopy. *Science* 267(5194):68-71.

76. Giessibl FJ (2019) The qPlus sensor, a powerful core for the atomic force microscope. *Rev. Sci. Instrum.* 90(1):011101.

77. Edwards H, Taylor L, Duncan W, & Melmed AJ (1997) Fast, high-resolution atomic force microscopy using a quartz tuning fork as actuator and sensor. *J. Appl. Phys.* 82(3):980-984.

78. Giessibl FJ (1998) High-speed force sensor for force microscopy and profilometry utilizing a quartz tuning fork. *Appl. Phys. Lett.* 73(26):3956-3958.

79. Rensen WHJ, van Hulst NF, Ruiter AGT, & West PE (1999) Atomic steps with tuning-fork-based noncontact atomic force microscopy. *Appl. Phys. Lett.* 75(11):1640-1642.

80. Giessibl FJ, Hembacher S, Bielefeldt H, & Mannhart J (2000) Subatomic Features on the Silicon (111)-(7×7) Surface Observed by Atomic Force Microscopy. *Science* 289(5478):422.

81. Giessibl FJ (2003) Advances in atomic force microscopy. *Rev. Mod. Phys.* 75(3):949-983.

82. Gross L (2011) Recent advances in submolecular resolution with scanning probe microscopy. *Nat. Chem.* 3(4):273-278.

83. Gross L, Mohn F, Moll N, Liljeroth P, & Meyer G (2009) The Chemical Structure of a Molecule Resolved by Atomic Force Microscopy. *Science* 325(5944):1110.

84. Giessibl FJ, Pielmeier F, Eguchi T, An T, & Hasegawa Y (2011) Comparison of force sensors for atomic force microscopy based on quartz tuning forks and length-extensional resonators. *Phys. Rev. B* 84(12):125409.

85. Wastl DS, Weymouth AJ, & Giessibl FJ (2013) Optimizing atomic resolution of force microscopy in ambient conditions. *Phys. Rev. B* 87(24).

86. Pürckhauer K, *et al.* (2018) Imaging in Biologically-Relevant Environments with AFM Using Stiff qPlus Sensors. *Sci. Rep.* 8(1):9330.

87. Herz M, Schiller C, Giessibl FJ, & Mannhart J (2005) Simultaneous current-, force-, and work-function measurement with atomic resolution. *Appl. Phys. Lett.* 86(15):153101.

88. Garcia V, *et al.* (2009) Giant tunnel electroresistance for non-destructive readout of ferroelectric states. *Nature* 460(7251):81-84.

89. Giessibl FJ (2000) Atomic resolution on Si(111)-(7×7) by noncontact atomic force microscopy with a force sensor based on a quartz tuning fork. *Appl. Phys. Lett.* 76(11):1470-1472.

90. Ooe H, *et al.* (2016) Amplitude dependence of image quality in atomically-resolved bimodal atomic force microscopy. *Appl. Phys. Lett.* 109(14):141603.

91. Ebeling D, *et al.* (2017) Chemical bond imaging using higher eigenmodes of tuning fork sensors in atomic force microscopy. *Appl. Phys. Lett.* 110(18):183102.

92. Albrecht TR, Grütter P, Horne D, & Rugar D (1991) Frequency modulation detection using high-Q cantilevers for enhanced force microscope sensitivity. *J. Appl. Phys.* 69(2):668.

93. Rabe U, Janser K, & Arnold W (1996) Vibrations of free and surface-coupled atomic force microscope cantilevers: Theory and experiment. *Rev. Sci. Instrum.* 67(9):3281-3293.

94. Hong S, *et al.* (2014) Charge gradient microscopy. *Proc. Natl. Acad. Sci. U.S.A.*:201324178.

95. Kalinin SV, *et al.* (2007) Towards local electromechanical probing of cellular and biomolecular systems in a liquid environment. *Nanotechnology* 18(42):424020.

96. Cordero-Edwards K, Domingo N, Abdollahi A, Sort J, & Catalan G (2017) Ferroelectrics as Smart Mechanical Materials. *Adv. Mater.* 29(37):1702210.

97. Shur VY, Akhmatkhanov AR, & Baturin IS (2015) Micro- and nano-domain engineering in lithium niobate. *Appl. Phys. Rev.* 2(4):040604.

98. Jungk T, Hoffmann Á, & Soergel E (2006) Quantitative analysis of ferroelectric domain imaging with piezoresponse force microscopy. *Appl. Phys. Lett.* 89(16):163507.

99. Wang H & Zeng K (2016) Domain structure, local surface potential distribution and relaxation of Pb(Zn$_{1/3}$Nb$_{2/3}$)O$_3$-9%PbTiO$_3$ (PZN-9%PT) single crystals. *J. Materiomics* 2(4):309-315.

100. Kalinin SV, *et al.* (2010) Direct evidence of mesoscopic dynamic heterogeneities at the surfaces of ergodic ferroelectric relaxors. *Phys. Rev. B* 81(6):064107.

101. Wang H & Zeng K (2020) Characterization of domain structure and imprint of Pb(Zn$_{1/3}$Nb$_{2/3}$)O$_3$-4.5%PbTiO$_3$ (PZN-4.5%PT) single crystals by using PFM and SS-PFM techniques. *Ceram. Int.* 46(4):4274-4279.

102. Meyer B & Vanderbilt D (2002) Ab initio study of ferroelectric domain walls in PbTiO$_3$. *Phys. Rev. B* 65(10):104111.

103. Kalinin SV, *et al.* (2006) Spatial resolution, information limit, and contrast transfer in piezoresponse force microscopy. *Nanotechnology* 17(14):3400-3411.

104. Jesse S, Baddorf AP, & Kalinin SV (2006) Switching spectroscopy piezoresponse force microscopy of ferroelectric materials. *Appl. Phys. Lett.* 88(6):062908.

105. Chen PJ & Montgomery ST (1980) A macroscopic theory for the existence of the hysteresis and butterfly loops in ferroelectricity. *Ferroelectrics* 23(1):199-207.

106. Morozovska AN, *et al.* (2007) Piezoresponse force spectroscopy of ferroelectric-semiconductor materials. *J. Appl. Phys.* 102(11):114108.

107. Guo HY, *et al.* (2002) Study of domain stability on $(Pb_{0.76}Ca_{0.24})TiO_3$ thin films using piezoresponse microscopy. *Appl. Phys. Lett.* 81(4):715-717.

108. Li Q, *et al.* (2015) Probing Local Bias-Induced Transitions Using Photothermal Excitation Contact Resonance Atomic Force Microscopy and Voltage Spectroscopy. *ACS Nano* 9(2):1848-1857.

109. Bonnell DA, Kalinin SV, Kholkin AL, & Gruverman A (2009) Piezoresponse Force Microscopy: A Window into Electromechanical Behavior at the Nanoscale. *MRS Bull.* 34(9):648-657.

110. Jesse S, Baddorf AP, & Kalinin SV (2006) Dynamic behaviour in piezoresponse force microscopy. *Nanotechnology* 17(6):1615-1628.

111. Grober RD, *et al.* (2000) Fundamental limits to force detection using quartz tuning forks. *Rev. Sci. Instrum.* 71(7):2776-2780.

112. Junquera J & Ghosez P (2003) Critical thickness for ferroelectricity in perovskite ultrathin films. *Nature* 422(6931):506-509.

113. Gao P, *et al.* (2017) Possible absence of critical thickness and size effect in ultrathin perovskite ferroelectric films. *Nat. Commun.* 8(1):15549.

114. Seidel J, *et al.* (2010) Domain Wall Conductivity in La-Doped $BiFeO_3$. *Phys. Rev. Lett.* 105(19):197603.

115. Seidel J, *et al.* (2009) Conduction at domain walls in oxide multiferroics. *Nat. Mater.* 8:229.

116. Rabe U & Arnold W (1994) Acoustic microscopy by atomic force microscopy. *Appl. Phys. Lett.* 64(12):1493-1495.

117. Dazzi A & Prater CB (2017) AFM-IR: Technology and Applications in Nanoscale Infrared Spectroscopy and Chemical Imaging. *Chem. Rev.* 117(7):5146-5173.

118. Lahiri B, Holland G, Aksyuk V, & Centrone A (2013) Nanoscale Imaging of Plasmonic Hot Spots and Dark Modes with the Photothermal-Induced Resonance Technique. *Nano Lett.* 13(7):3218-3224.

119. Chen QN, *et al.* (2013) High sensitivity piezomagnetic force microscopy for quantitative probing of magnetic materials at the nanoscale. *Nanoscale* 5(13):5747-5751.

120. Varesi J & Majumdar A (1998) Scanning Joule expansion microscopy at nanometer scales. *Appl. Phys. Lett.* 72(1):37-39.

121. Kalinin SV, Jesse S, Tselev A, Baddorf AP, & Balke N (2011) The Role of Electrochemical Phenomena in Scanning Probe Microscopy of Ferroelectric Thin Films. *ACS Nano* 5(7):5683-5691.

## Acknowledgments

This research is supported by Ministry of Education (MoE) Singapore through National University of Singapore (NUS) under the Academic Research Fund (AcRF) of R-265-000-596-112 and R-265-000-A27-114. Q. Z. would like to acknowledge the scholarship support from MoE Singapore through NUS under AcRF of R-265-100-596-112 and Department of Mechanical Engineering, NUS. D. C. thanks the financial support from the National Natural Science Foundation of China (Grant Nos. U1832104 and 91963102), Guangdong Science and Technology (Grant No. 2019A050510036) and Guangdong Provincial Key Laboratory of Optical Information Materials and Technology (No. 2017B030301007). We would also like to thank Prof. Li Lu (NUS) for great support of the SPA400 SPM in this study.

## Author Contributions

Q. Z. designed and constructed the model NC-HEsFM system, conducted all the measurements and calculations, analyzed the results, and wrote the manuscript. H. W. prepared the PZN-9%PT sample. Q. H. and Z. F. prepared the PZT film. C. L. and D. C. prepared the BFO film. Q. Z. and Y. C. performed the finite element simulations. K. Z. led the project and directed all experimental research as well as reviewed the manuscript. Q. Z. and K. Z. conceived the original idea. All authors commented on the manuscript and approved its submission.

## Competing Interests

The authors declare that they have no competing interests.

## Data and Materials Availability

All data needed to evaluate the conclusions in the paper are present in the paper and/or the Supplementary Materials. Additional data related to this paper may be requested from the authors.

# Figures

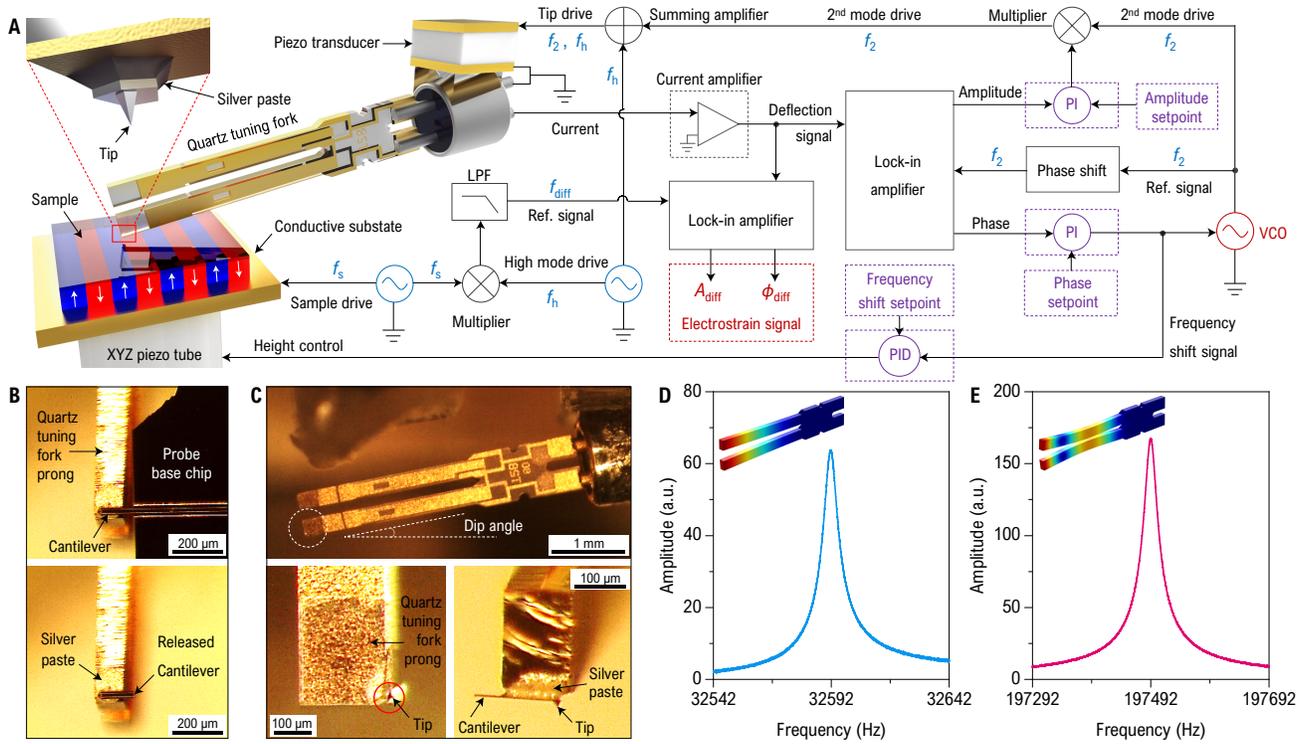

**Fig. 1. Experimental set-up of NC-HEsFM.** (A) Schematic diagram of the NC-HEsFM system. The QTF is mechanically excited by the 2$^{nd}$ and high mode drives with frequencies of $f_2$ and $f_h$, respectively, while the sample is electrically excited by the sample drive with frequency of $f_s$. The difference-frequency electrostrain signal generated from the nonlinear tip-sample interaction has a frequency of $f_{diff} = f_s - f_h$. (B) Optical photographs about transferring the nano-tip from conventional AFM probe to QTF prong. (C) Optical photographs of the finished QTF force sensor with nano-tip installed at the end of the QTF prong. (D) The free resonance curves of the 1$^{st}$ and (E) the 2$^{nd}$ flexural mode measured from a finished QTF sensor in air. Insets show the corresponding mode shapes obtained from the finite element simulation.

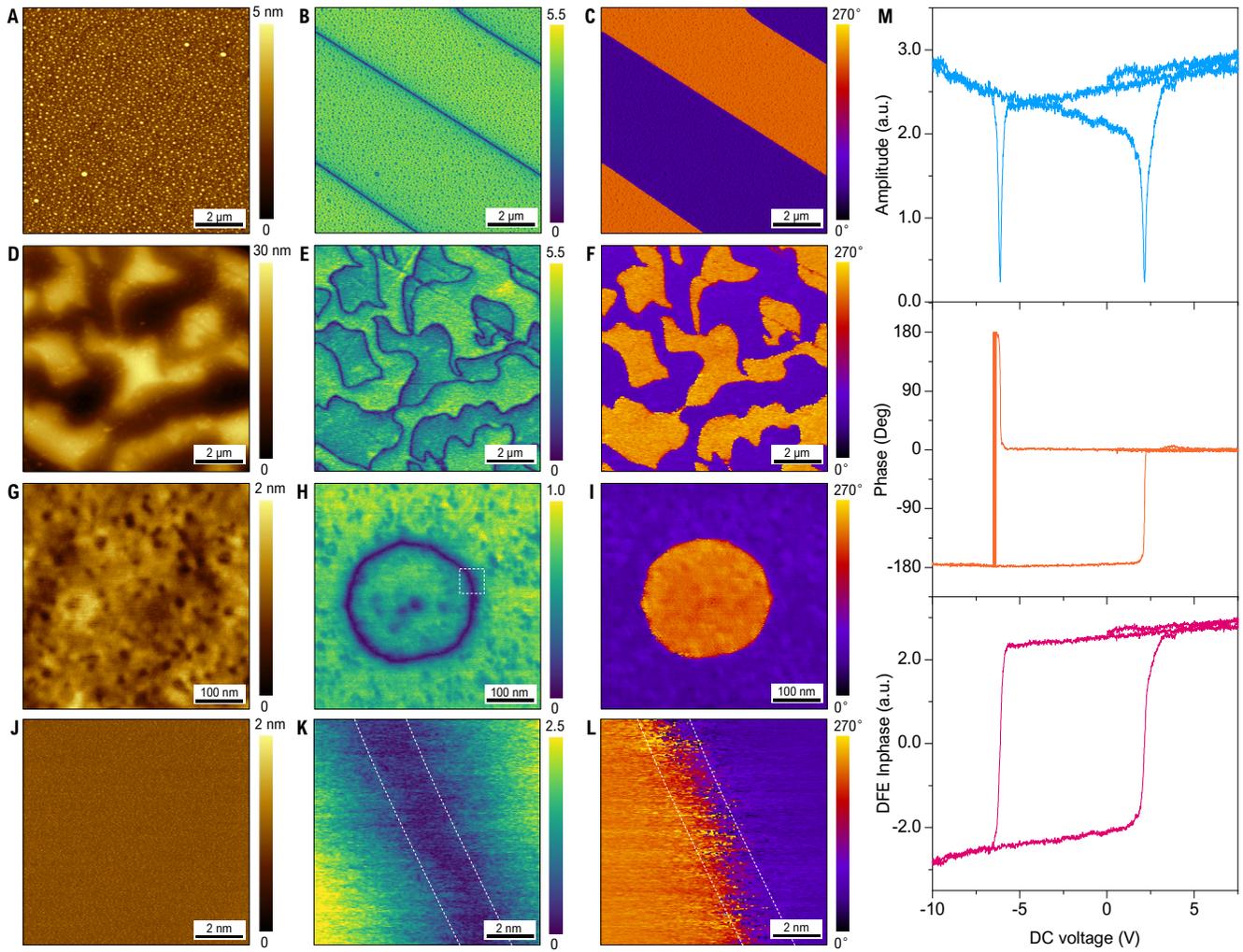

**Fig. 2. NC-HEsFM ferroelectric characterization on standard ferroelectric materials.** (A) Topography, (B) DFE amplitude and (C) DFE phase image of the PPLN sample. (D) Topography, (E) DFE amplitude and (F) DFE phase image of the PZN-9%PT sample. (G) Topography, (H) DFE amplitude and (I) DFE phase image of an artificial domain created on PZT film. (J) Topography, (K) DFE amplitude and (L) DFE phase image obtained around the PZT domain wall (indicated by the white box in (H)). (M) Amplitude (top), phase (middle) and inphase (bottom) hysteresis loops of the PZT film. Measurement conditions: (A-C) $f_h$ = 1459.475 kHz, $f_s$ = 1492.008 kHz, $\Delta f_{sp}$ = +20 Hz, $A_{sp}$ = 10 mV; (D-F) $f_h$ = 947.110 kHz, $f_s$ = 979.475 kHz, $\Delta f_{sp}$ = +35 Hz, $A_{sp}$ = 10 mV; (G-I) $f_h$ = 955.658 kHz, $f_s$ = 988.305 kHz, $\Delta f_{sp}$ = +25 Hz, $A_{sp}$ = 10 mV; (J-L) $f_h$ = 955.658 kHz and $f_s$ = 988.395 kHz, $\Delta f_{sp}$ = +40 Hz, $A_{sp}$ = 2 mV; (M) $f_h$ = 953.909 kHz, $f_s$ = 986.424 kHz, $\Delta f_{sp}$ = +50 Hz, $A_{sp}$ = 5 mV.

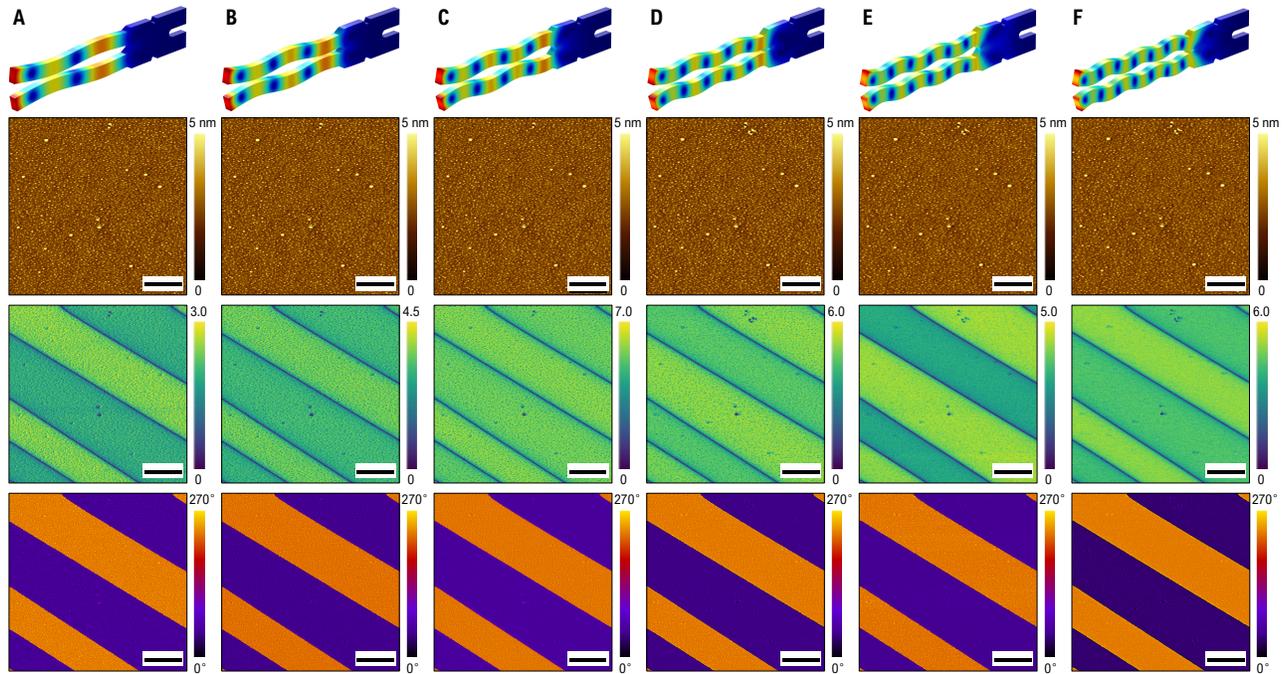

**Fig. 3. NC-HEsFM ferroelectric characterization with different eigenmodes.** (A-F) *In-situ* NC-HEsFM measurement results of PPLN sample by using the 3$^{rd}$ to the 8$^{th}$ flexural mode of the QTF, respectively. The 1$^{st}$ raw to 4$^{th}$ raw respectively show the finite element simulated mode shapes, topography images, DFE amplitude and DFE phase images. Measurement conditions: (A) $f_h$ = 522.508 kHz, $f_s$ = 555.107 kHz; (B) $f_h$ = 953.954 kHz, $f_s$ = 986.553 kHz; (C) $f_h$ = 1459.566 kHz, $f_s$ = 1492.165 kHz; (D) $f_h$ = 2015.098 kHz, $f_s$ = 2047.697 kHz; (E) $f_h$ = 2594.125 kHz, $f_s$ = 2626.724 kHz; (F) $f_h$ = 3195.517 kHz, $f_s$ = 3228.116 kHz; (A-F) $f_{diff}$ = 32.599 kHz, $\Delta f_{sp}$ = +25 Hz, $A_{sp}$ = 10 mV. Scale bar in (A-F) is 3 μm.

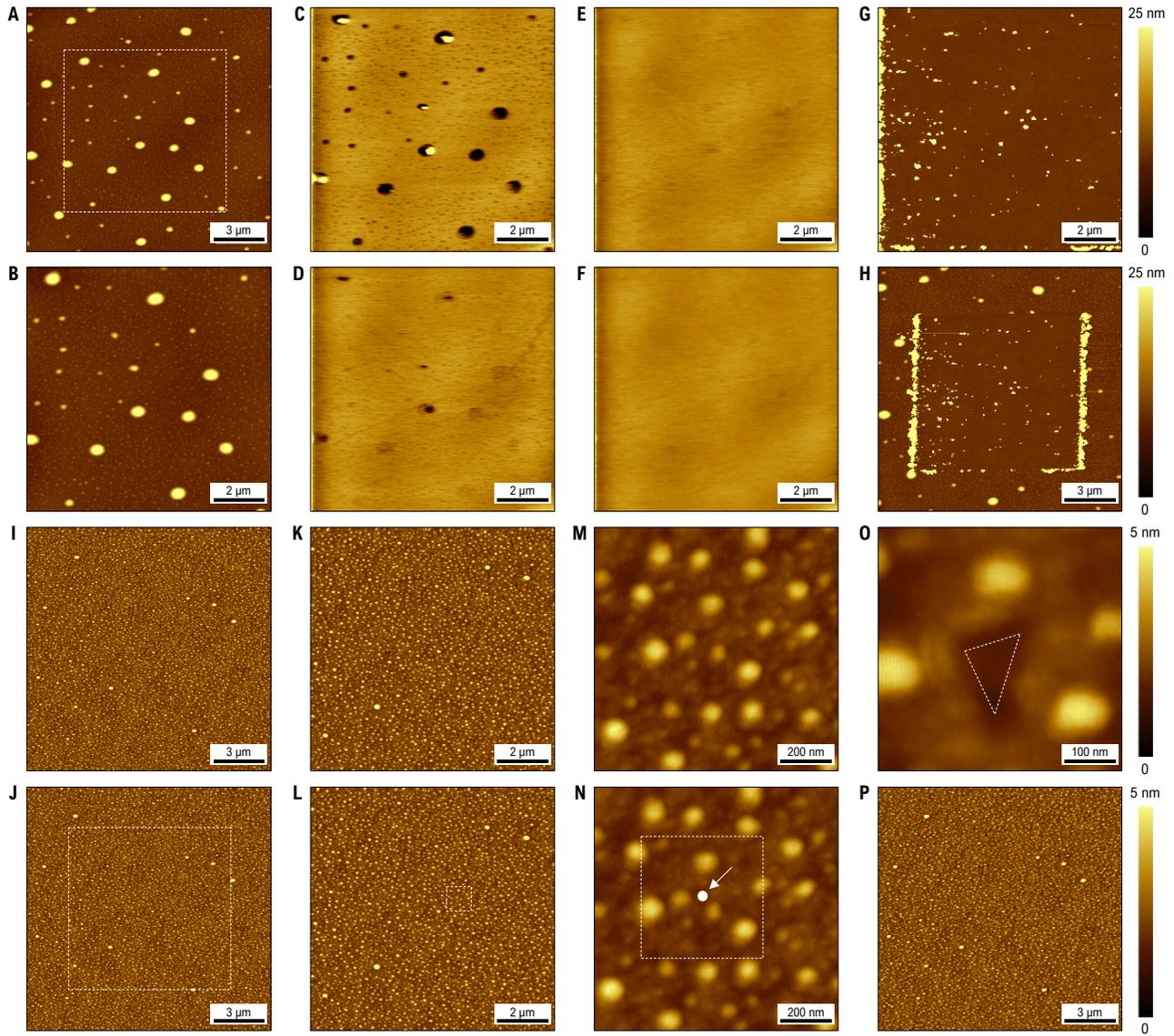

**Fig. 4. Comparison of the surface modification situations between the conventional PFM and NC-HEsFM measurements.** (A) Initial and (B) zoomed-in topography images measured by tapping-mode AFM. (C-F) The zoomed-in topography images obtained from the 1st to the 4th PFM scanning, respectively. (G) The topography image measured after the PFM scanning using tapping-mode AFM, and the scanning area of (C-G) is indicated by the white box in (A). (H) Final topography images measured by tapping AFM after the PFM scans. (I) Initial topography images measured by the 1st and (J) the 2nd NC-HEsFM scanning. (K) The topography of the 1st and (L) the 2nd zoomed-in NC-HEsFM scanning, the scanning area is indicated by the white box in (J). (M) The topography of the 3rd and (N) the 4th zoomed-in NC-HEsFM scans, the scanning area is indicated by the white box in (L). (O) Topography image of *in-situ* NC-HEsFM scanning after the weak indentation, the scanning area is indicated by the white box in (N). (P) Final topography image measured by NC-HEsFM after zoomed-in scanning. Measurement conditions: (I-P) $f_h$ = 1459.475 kHz, $f_s$ = 1492.008 kHz, $\Delta f_{sp}$ = +20 Hz and $A_{sp}$ = 10 mV.

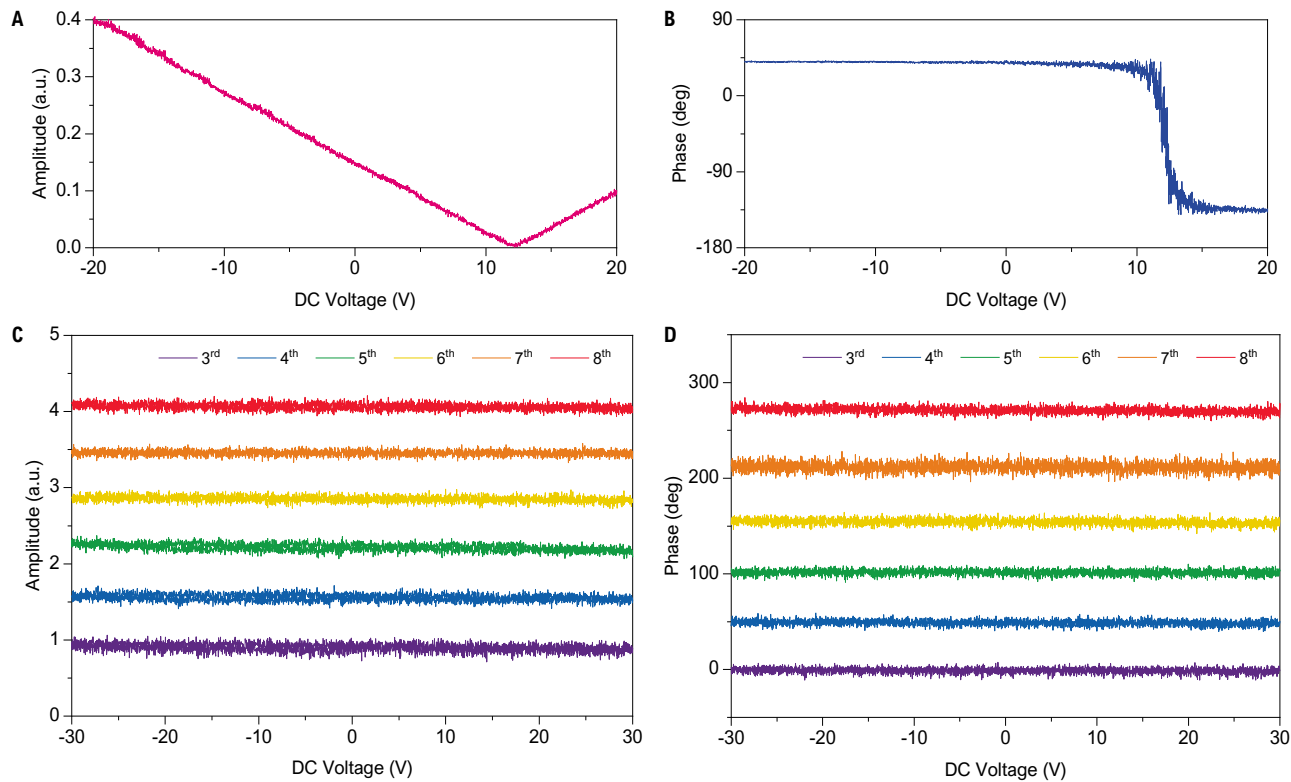

**Fig. 5. Electrostatic force contribution in conventional PFM and NC-HEsFM.** (A) The PFM amplitude and (B) phase as a function of DC voltage measured by PFM on PPLN sample. (C) The DFE amplitude and (D) DFE phase as a function of DC voltage measured by NC-HEsFM on PPLN sample using the 3$^{rd}$ to the 8$^{th}$ flexural mode. Note that all the spectrums in (C) and (D) are offset for clarity. Measurement conditions: (C, D) $f_{\text{diff}}$ = 32.634 kHz, $\Delta f_{\text{sp}}$ = +40 Hz, $A_{\text{sp}}$ = 5 mV.

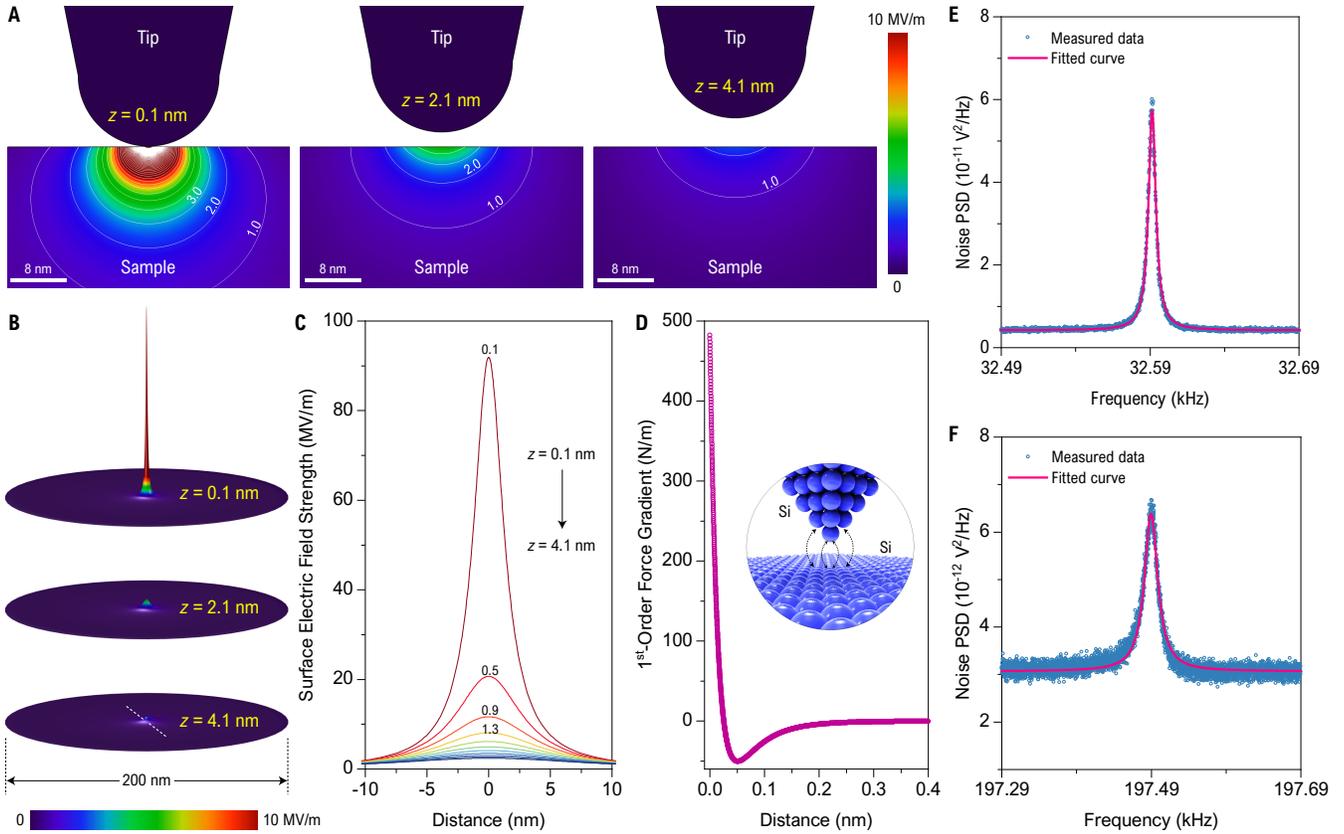

**Fig. 6. Mechanism of achieving effective electrostrain measurement.** (A) Finite element simulated cross-sectional and (B) surface electric field distribution of a 1 V biased tip on sample under different tip-sample distance ($z$), the white arc lines shown in (A) are electric field strength contour lines with interval of 1 MV/m. (C) Line profiles of the surface electric field strength extracted from (B) under a series of $z$. (D) The calculated 1st-order force gradient as a function of $z$ for ideal tip-sample (Si tip on Si surface) interaction. (E) The experimental measured and fitted voltage noise PSD spectrums of the 1st and (F) the 2nd flexural mode.

Supplementary Materials for

# Nanoscale Non-Destructive Ferroelectric Characterization with Non-Contact Heterodyne Electrostrain Force Microscopy

Qibin Zeng, Qicheng Huang, Hongli Wang, Caiwen Li, Zhen Fan, Deyang Chen, Yuan Cheng, Kaiyang Zeng*

*Corresponding author: Kaiyang Zeng

**Email:** mpezk@nus.edu.sg

**ORCID:** Kaiyang Zeng 0000-0002-3348-0018

**This PDF file includes:**

    Supplementary Text S1 to S8
    Figs. S1 to S10
    Table S1
    References

## S1. Capacitance Coupling of the QTF

Due to the open-structure design of QTF sensor, there exists a significant capacitance coupling between the sample and QTF sensor, which will lead to a large co-frequency interference in the final output signal when AC drive is applied to sample. Fig. S1$A,B$ show the frequency spectrums of the current amplifier output signal, which are acquired when NC-HEsFM tip is located ~100 μm from the PPLN surface. Obviously, with applying sample drive (32.569 kHz, 2 $V_{pp}$), a significant co-frequency interference can be observed in the frequency spectrum of the output signal (Fig. S1$B$). Therefore, if the conventional PFM's homodyne detection is used for QTF-based electrostrain measurement, the tiny electrostrain signal will be totally masked by the strong capacitance coupling signal, causing the demodulation of this target signal almost impossible.

## S2. Difference-Frequency Background Signal

During NC-HEsFM measurement, there are 3 frequency components exist in the frequency spectrum of the current amplifier output signal, including the resonance signals of the 2$^{nd}$ ($f_2$) and high ($f_h$) flexural modes as well as a capacitance coupling signal ($f_s$). Due to the nonlinear effect of the current amplifying circuit, the frequency mixing can occur between any two frequency components. Since the difference-frequency electrostrain (DFE) signal has the frequency of $f_{diff} = f_s - f_h$, the frequency mixing between high mode and capacitance coupling signals can directly affect the DFE signal as this process will generate a difference-frequency background signal which has the same frequency with DFE signal. To verify the existence of the difference-frequency background signal, a control experiment has been implemented, in which the frequency spectrums of the output signal have been acquired with and without applying sample and high mode drives and the results are displayed in Fig. S4. Fig. S4$A$ shows the frequency spectrum with only the 2$^{nd}$ mode drive is applied (the tip is withdrawn 20 μm from the PPLN surface), and a zoomed-in view of a narrow flat band is shown in Fig. S4$C$. It is clear that, within the 34 to 35 kHz region, there is no frequency component can be observed. After applying the high mode ($f_h$= 1458.773 kHz, 0.5 $V_{pp}$) and sample ($f_s = f_h$ + 34.5 kHz = 1493.273 kHz, 4 $V_{pp}$) drives, the frequency spectrums shown in Fig. S4$A,C$ change to Fig. S4$B,D$ respectively. Comparing Fig. S4$C$ and S4$D$, an obvious spectral peak at 34.5 kHz emerges when sample and high mode drives are applied. Since the difference frequency is intentionally set to 34.5 kHz here, the peak shown in Fig. S4$D$ exactly indicates the generation of difference-frequency background signal. Typically, the difference-frequency background signal has an amplitude of ~100 μV, while the target DFE signal has an amplitude ranges from a few hundred to thousand μV, thereby sometimes the difference-frequency background signal can cause a pronounced influence to the DFE signal, especially when the excited sample electrostrain is very small. Since the difference-frequency background component is generated during the current amplifying process, it can be minimized by modifying the current amplifying circuit to either reduce the non-linear effect or suppress the high-frequency inputs. On the other hand, it is found that the incomplete electromagnetic shielding of the piezo transducer can cause a remarkable capacitance coupling between the transducer and QTF, which will additionally induce a coupling signal (with the same frequency of high mode drive) in the current amplifier and finally cause the difference-frequency background component get further enhanced. Therefore, a good electromagnetic shielding of the piezo transducer is also important for minimizing the difference-frequency background signal.

## S3. Ferroelectric Domain Characterization by Conventional PFM

Conventional contact-mode PFM scanning has been performed on PPLN sample and some typical results are displayed in Fig. S5. Fig. S5*A*,*B* and S5*C*,*D* show the PFM images obtained by continuously *in-situ* scanning at two areas, respectively. It is clear that at both the two areas, the surface topographies have been changed by the continuous contact-mode PFM scanning, and at the same time, the PFM amplitude and phase images acquired in the third scanning (Fig. S5*B*,*D*) manifest much better than that of the first scanning (Fig. S5*A*,*C*).

## S4. Non-Destructive Measurement and Stability Test of NC-HEsFM

To further verify the non-destructive measurement capability of NC-HEsFM, repeated NC-HEsFM scanning has been implemented on PPLN at several areas with large protrusion structures, and the results are shown in Fig. S6. From Fig. S6, it can be seen that both the large protrusions and fine structures on PPLN surface can be well protected during the NC-HEsFM scanning, which unambiguously demonstrates the non-destructive measurement capability of NC-HEsFM. In addition, a much stricter and longer experiment which includes 16 times *in-situ* scanning (last for ~9 hours, no withdrawing the tip) on PLLN has been conducted to test the stability and reproducibility of NC-HEsFM. Fig. S7 shows the topographies obtained from the *in-situ* scanning under the stimulation of seven different eigenmodes. Comparing the initial (Fig. S7*A*) and final (Fig. S7*P*) topography images, almost no surface modifications can be observed after the 14 times zoomed-in scanning. Meanwhile, Fig. S7*B-O* show that the obtained topography results are highly reproducible and have no noticeable relationship with the high mode, implying that the NC-HEsFM measurement can be highly stable during the continuous scanning as well as the operation parameter adjustment.

## S5. Tip-Sample Interaction Force

For simplification, here a well-studied Si-Si (Si tip on Si surface) interaction model (*1*) is used to demonstrate the relationship between tip-sample interaction force and tip-sample distance. Considering both the long-range van der Waals potential and short-range Lennard-Jones potential, the total tip-sample force $F_{ts}(z)$ for an idealized pyramidal Si tip (bounded by (111) planes with full tip angle of 70.5°) on a flat Si surface can be calculated explicitly by (*1*)

$$F_{ts}(z) = -1 \times 10^{-10} \frac{\sigma}{z} - 1.89 \times 10^{-8} \left( \frac{\sigma^7}{z^7} - \frac{\sigma^{13}}{z^{13}} \right) \tag{S1}$$

where $\sigma = 0.235$ nm is the nearest-neighbor distance of Si atom (*1*). Taking the derivative of Equation (S1), the 1$^{st}$-order (using the negative form here) and the 2$^{nd}$-order force gradient can be obtained as

$$-F_{ts}'(z) = -1 \times 10^{-10} \frac{\sigma}{z^2} - 1.89 \times 10^{-8} \left( 7\frac{\sigma^7}{z^8} - 13\frac{\sigma^{13}}{z^{14}} \right) \tag{S2}$$

$$F_{ts}''(z) = -2 \times 10^{-10} \frac{\sigma}{z^3} + 1.89 \times 10^{-8} \left( -56\frac{\sigma^7}{z^9} + 182\frac{\sigma^{13}}{z^{15}} \right) \tag{S3}$$

According to Equation (S1-S3), the tip-sample force and force gradients can be plotted as a function of tip-sample distance, which are all shown in Fig. S8.

## S6. Finite Element Simulation

The modal analysis of the QTF and the electric field distribution simulation are finished by using the finite element simulation software package COMSOL Multiphysics® 5.3 (COMSOL Inc.). Fig. S9*A,B* show the 3D model and mesh of the QTF used for modal analysis. The length, width, and thickness of the QTF prong are measured to be 2340 μm, 130 μm and 225 μm, respectively, which are consistent with other studies (*2-4*). In the simulation, the density, Young's modulus and Poisson's ratio of the quartz are set as 2648 kg/m$^3$, 82 GPa and 0.17, respectively (*4, 5*). Table S1 shows the calculated as well as the experimentally measured eigenfrequencies of the 1$^{st}$ to the 8$^{th}$ flexural mode (antisymmetric vibration) of the QTF, in which the simulation well matches with the experiment.

For the simulation of electric field distribution, a 2D axisymmetric model is used for the purpose of simplification, and the 2D model and mesh are displayed in Fig. S9*C,D*. After the 2D simulation is completed, a 3D data set (schematically shown in Fig. S9*E*) of electric field distribution can be generated by performing 2D revolution. In the simulation, the tip radius, sample thickness and sample radius are set to 10 nm, 200 nm and 200 nm, respectively. The sample is modeled as an isotropic dielectric with a relative permittivity of 100. The tip bias is set to 1 V while the bottom surface of the sample is grounded. During the simulation, the tip-sample distance is changed from 0.1 nm to 4.1 nm with the step of 0.4 nm.

**S7. Voltage Noise PSD Model**

When the two prongs of the QTF are excited by the thermal noise force, the total current generated from the two prongs can be expressed as

$$I(t) = c[x_1(t) - x_2(t)] \tag{S4}$$

in which $x_1(t)$ and $x_2(t)$ are the thermal noise force-induced displacements of the two prongs, and $c$ is a piezoelectric coupling constant. As the QTF current is amplified by the current amplifier with a gain of $R_{gain}$, the output voltage of the amplifier is given by

$$V(t) = R_{gain} \cdot I(t) = cR_{gain}[x_1(t) - x_2(t)] \tag{S5}$$

We define the voltage deflection sensitivity as $\alpha = cR_{gain}$. According to the Wiener-Khinchin theorem, the power spectral density (PSD) of the output voltage is calculated by

$$S_{VV}(\omega) = \int_{-\infty}^{+\infty} R_{VV}(\tau) e^{-i\omega\tau} d\tau \tag{S6}$$

where $R_{VV}(\tau)$ is the autocorrelation function of $V(t)$. As the two QTF prongs are identical, under the approximation that they are only very weakly coupled during thermal motion (i.e., their thermal motion is approximately uncorrelated) (6), the PSD of the output voltage can be expressed as

$$S_{VV}(\omega) = 2\alpha^2 \int_{-\infty}^{+\infty} R_{x_1 x_1}(\tau) e^{-i\omega\tau} d\tau = 2\alpha^2 S_{x_1 x_1}(\omega) \tag{S7}$$

in which $R_{x_1 x_1}(\tau)$ and $S_{x_1 x_1}(\omega)$ are the autocorrelation function and PSD of $x_1(t)$, respectively. Using the damped harmonic oscillation model, the PSD of $x_1(t)$ can be calculated by (7)

$$S_{x_1 x_1}(\omega) = \frac{1/m^2}{\left(\omega_0^2 - \omega^2\right)^2 + \left(\omega_0 \omega / Q\right)^2} S_{NN}(\omega) \tag{S8}$$

where $m$ and $\omega_0$ are equivalent mass and resonance frequency of the QTF prong, while $S_{NN}(\omega)$ is the PSD of the thermomechanical noise force. According to the fluctuation-dissipation theorem, $S_{NN}(\omega)$ is given by (8, 9)

$$S_{NN}(\omega) = 4k_B T \xi \tag{S9}$$

where $k_B$, $T$ and $\xi$ are Boltzmann constant, temperature and damping coefficient. Combing Equation (S7-S9), the PSD of $x_1(t)$

can be obtained as

$$S_{x_1 x_1}(\omega) = \frac{4k_B T \xi / m^2}{\left(\omega_0^2 - \omega^2\right)^2 + \left(\omega_0 \omega / Q\right)^2} = \frac{4k_B T \omega_0^3 / (k_0 Q)}{\left(\omega_0^2 - \omega^2\right)^2 + \left(\omega_0 \omega / Q\right)^2} \tag{S10}$$

Substituting Equation (S10) into (S7) and using the temporal frequency expression, the PSD of the thermal noise-induced output voltage can be obtained by

$$S_{VV}(f) = \alpha^2 \frac{4k_B T f_0^3 / (\pi k_0 Q)}{\left(f_0^2 - f^2\right)^2 + \left(f_0 f / Q\right)^2} \tag{S11}$$

Therefore, the voltage deflection sensitivity $\alpha$ can be determined by fitting the measured voltage noise PSD spectrum using Equation (S11).

## S8. Measurement of High-Order Electrostrain

The NC-HEsFM developed here can be easily modified to measure the high-order electrostrains, such as the $2^{nd}$-order electrostrictive strain. Conventional PFM-based method has been extensively used to study the electrostriction (*10-13*), however this method is subject to many similar limitations, such as the contact-mode operation, the influences of the $2^{nd}$-harmonic electrostatic force (*14*) and the Joule thermal expansion (*13*). With the advantages of QTF sensor and heterodyne detection, the measurement of the electrostriction can get substantial improvements by using NC-HEsFM-based method with non-contact manner (*15*). Firstly, the $2^{nd}$-harmonic electrostatic force contribution can be significantly minimized in NC-HEsFM-based method due to the ultra-high stiffness of QTF and heterodyne detection, and then the non-contact measurement can help to reduce the tip-sample current thus the Joule thermal expansion can be largely suppressed. As the electrostriction is a quadratic effect, the target electrostrictive strain will locate at the $2^{nd}$-harmonic of the sample drive. Therefore, to measure the electrostrictive strain with NC-HEsFM, the difference frequency $f_{diff}$ should be changed to $2f_s - f_h$. In a similar fashion, the $3^{rd}$-order electrostrain can be measured by setting $f_{diff} = 3f_s - f_h$, and more generally, the $N^{th}$-order ($N$ = 1, 2, 3…) electrostrain can be measured by setting $f_{diff} = Nf_s - f_h$. If the clocks of signal generator and lock-in amplifier can be synchronized, then detecting these high-order electrostrains by NC-HEsFM will be quite straightforward. Fig. S10 shows a modified NC-HEsFM set-up for the measurement of $N^{th}$-order electrostrain (*15*), where the reference signal for demodulating the electrostrain signal is provided internally *via* the clock synchronization between signal generator and lock-in amplifier.

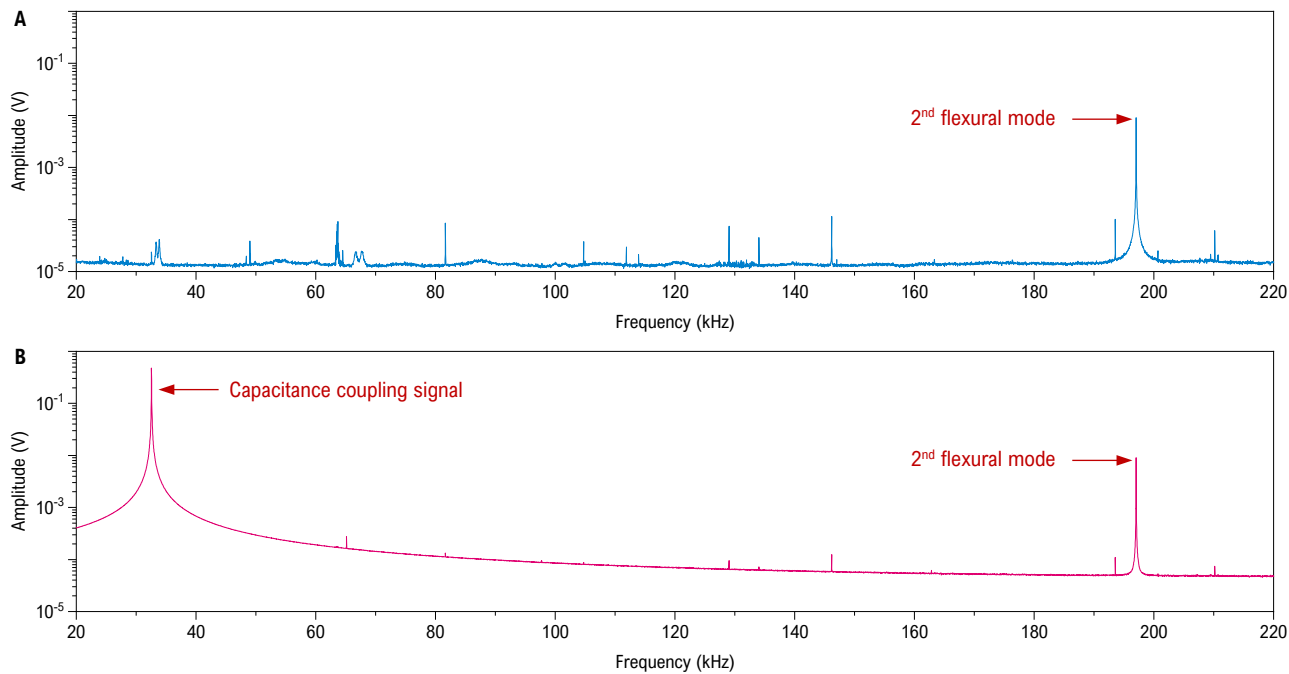

**Fig. S1. Capacitance coupling of the QTF sensor.** (A) The frequency spectrum of the current amplifier output signal without and (B) with applying sample drive.

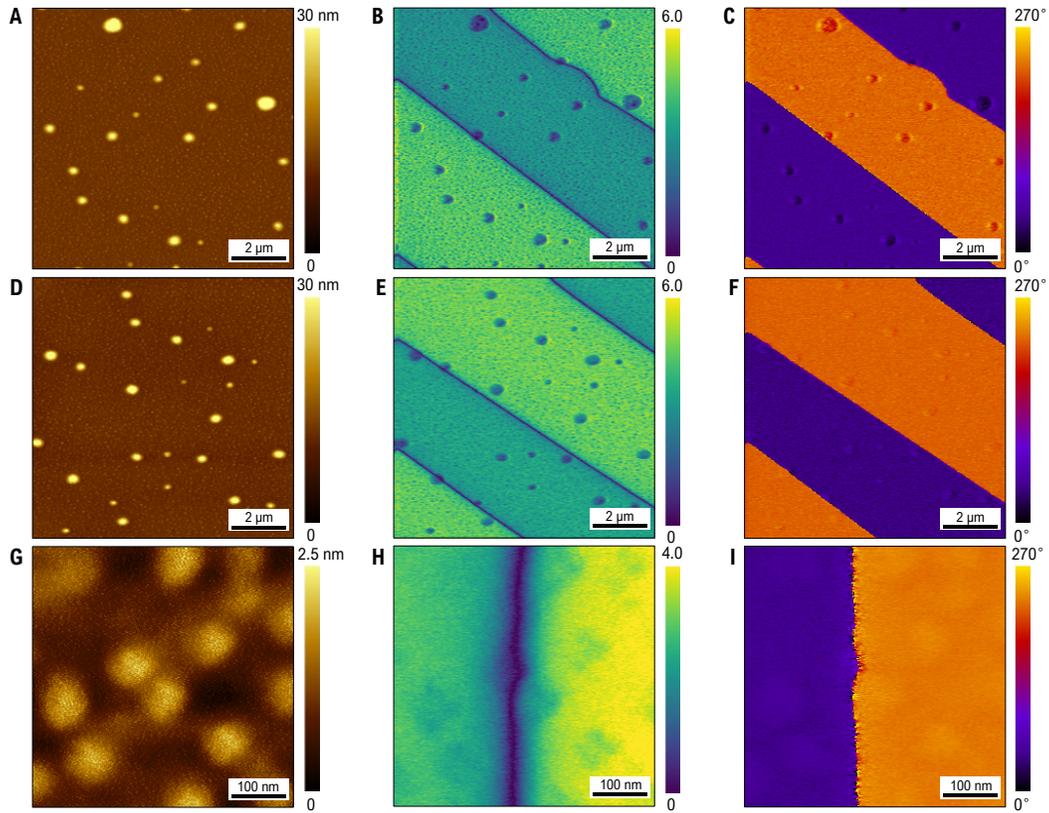

**Fig. S2. Supplementary data of NC-HEsFM measurement on PPLN.** (A, D, G) Topography, (B, E, H) DFE amplitude and (C, F, I) DFE phase images of PPLN. Measurement conditions: (A-C) $f_h$ = 953.725 kHz, $f_s$ = 986.523 kHz, $\Delta f_{sp}$ = +20 Hz, $A_{sp}$ = 15 mV; (D-I) $f_h$ = 954.438 kHz, $f_s$ = 987.015 kHz, $\Delta f_{sp}$ = +20 Hz, $A_{sp}$ = 15 mV.

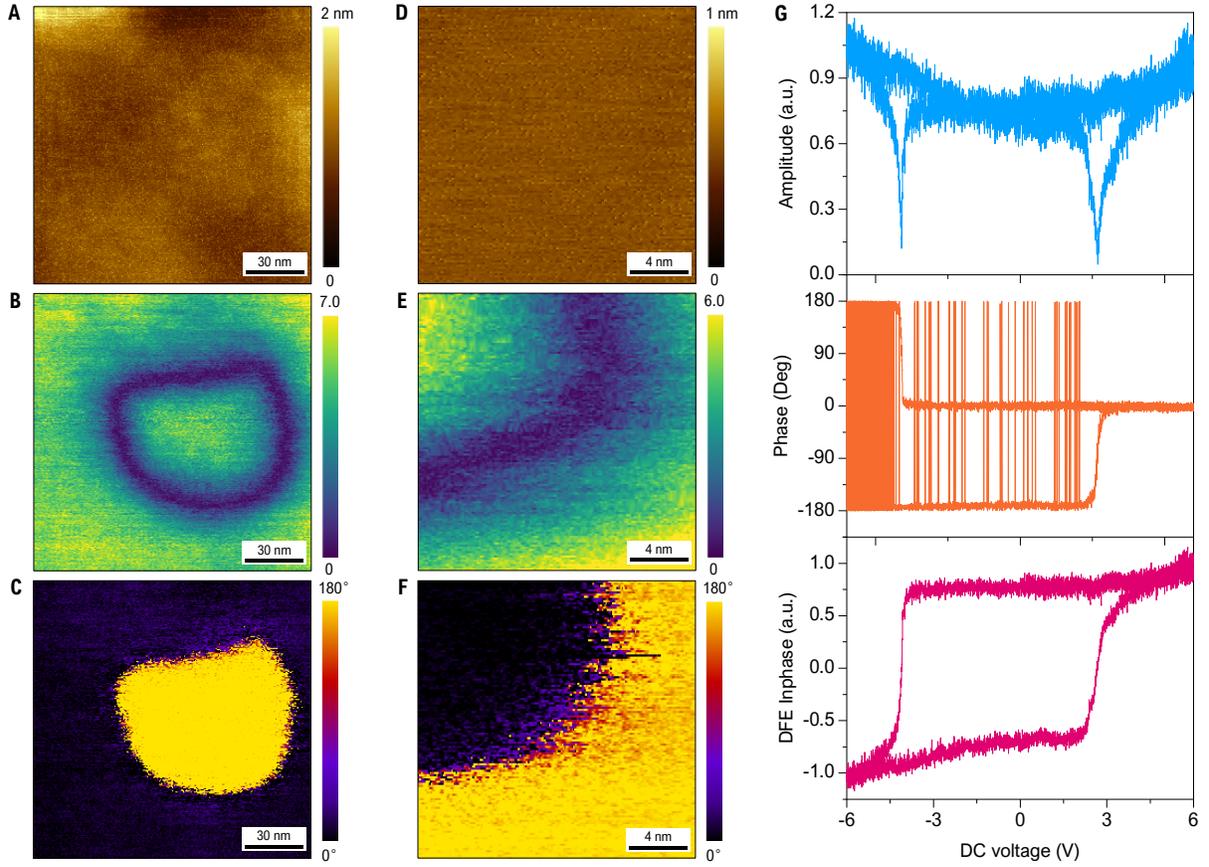

**Fig. S3. NC-HEsFM ferroelectric characterization on BFO film.** (A) Topography, (B) DFE amplitude and (C) DFE phase image of an artificial domain on BFO film. (D) Topography, (E) DFE amplitude and (F) DFE phase image of an artificial domain wall. (G) Amplitude (top), phase (middle) and inphase (bottom) hysteresis loops of the BFO film. Measurement conditions: (A-C) $f_h$ = 953.731 kHz, $f_s$ = 986.305 kHz, $\Delta f_{sp}$ = +30 Hz, $A_{sp}$ = 10 mV; (D-F) $f_h$ = 953.731 kHz, $f_s$ = 987.235 kHz, $\Delta f_{sp}$ = +40 Hz, $A_{sp}$ = 3 mV; (G) $f_h$ = 953.731 kHz, $f_s$ = 987.334 kHz, $\Delta f_{sp}$ = +50 Hz, $A_{sp}$ = 5 mV.

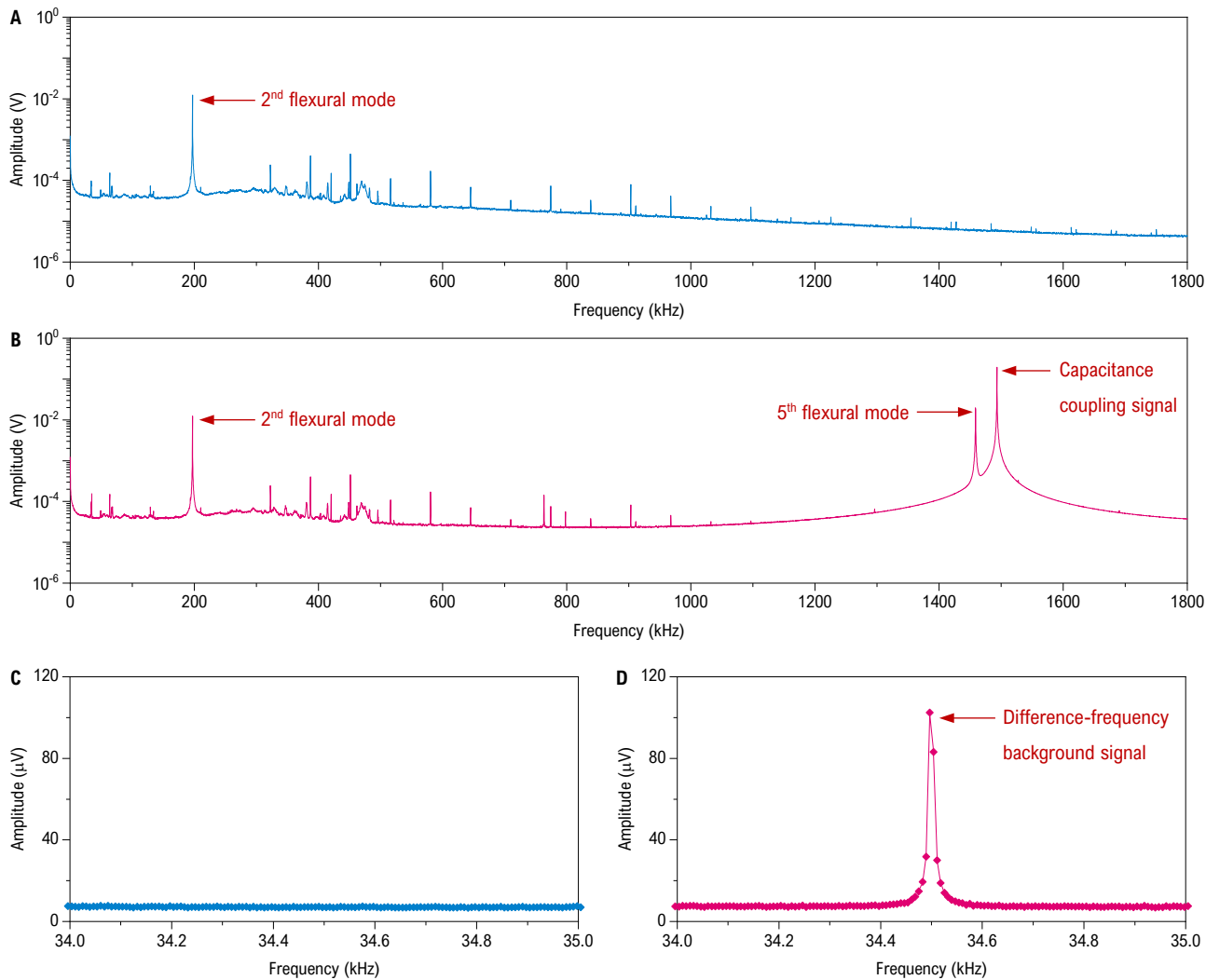

**Fig. S4. Difference-frequency background signal of the current amplifier.** (A) The frequency spectrum of the current amplifier output signal with only the 2nd flexural mode is excited. (B) The frequency spectrum of the current amplifier output signal with applying both sample and high mode drives. (C, D) A zoomed-in view of the frequency spectrum of (A) and (B), respectively.

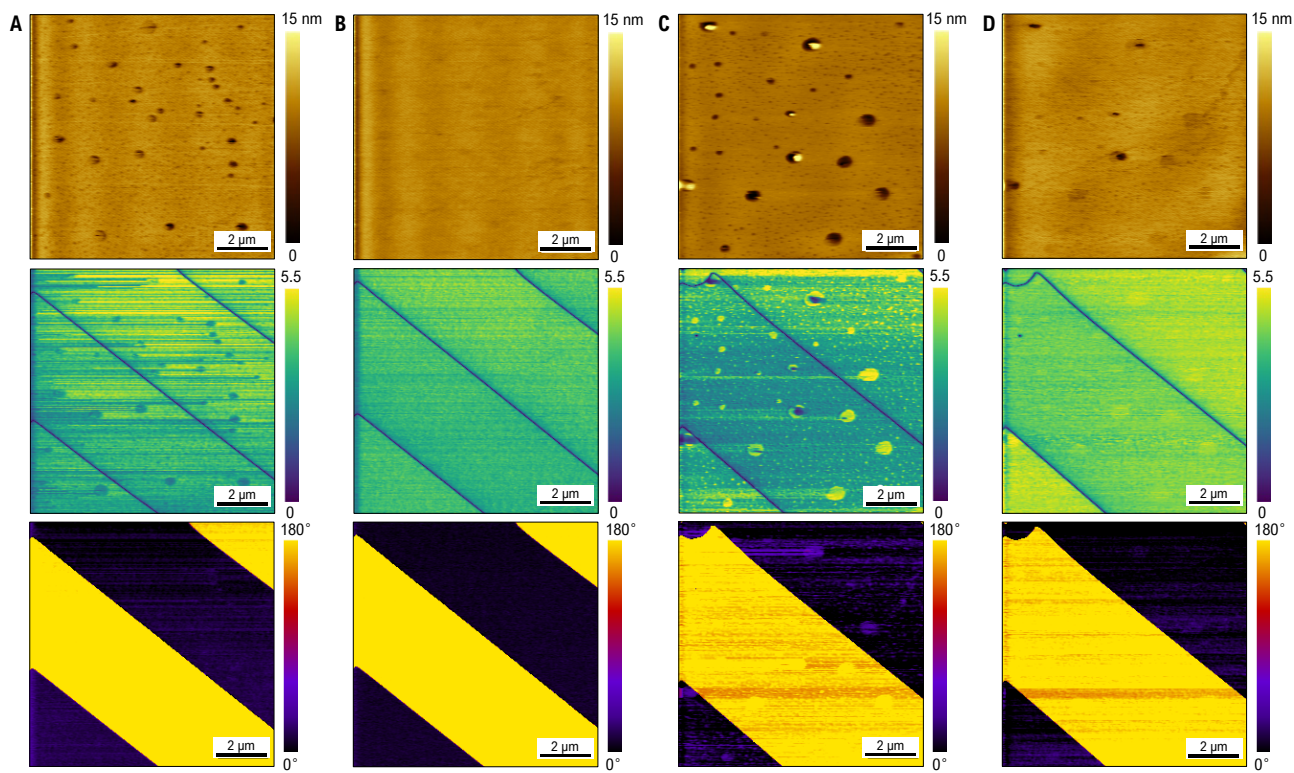

**Fig. S5. Ferroelectric characterization of PPLN by conventional PFM.** (A, C) Ferroelectric domain images obtained by the first and (B, D) the third PFM scanning. The 1st raw to 3rd raw respectively show the topography images, PFM amplitude and PFM phase images.

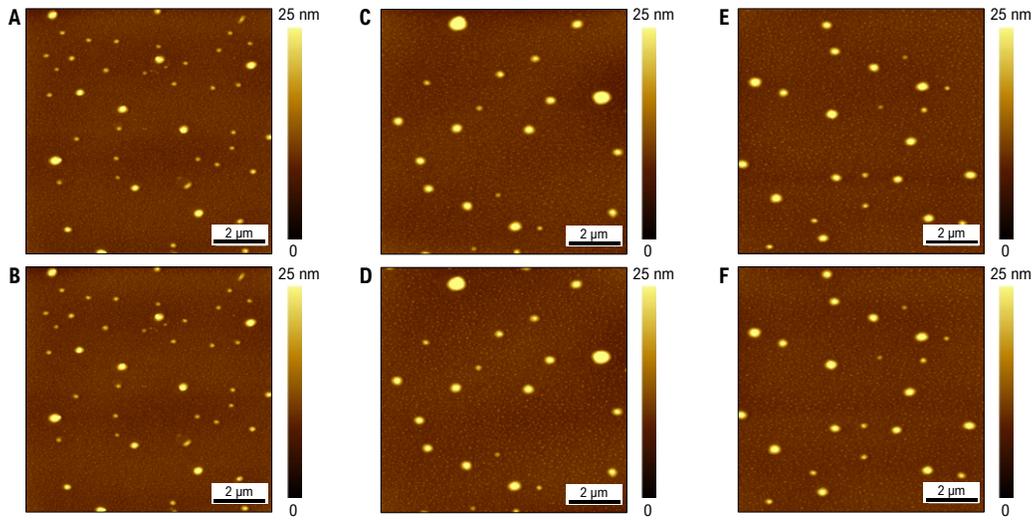

**Fig. S6. Non-destructive scanning on PPLN by NC-HEsFM.** (A, C, E) The topography image obtained by the first and (B, D, F) the respectively second NC-HEsFM scanning on PPLN surface with large protrusions. Measurement conditions: (A, B) $f_h$ = 953.723 kHz, $f_s$ = 987.014 kHz; (C, D) $f_h$ = 953.725 kHz, $f_s$ = 986.523 kHz; (E, F) $f_h$ = 954.438 kHz, $f_s$ = 987.015 kHz; (A-F) $\Delta f_{sp}$ = +20 Hz, $A_{sp}$ = 15 mV.

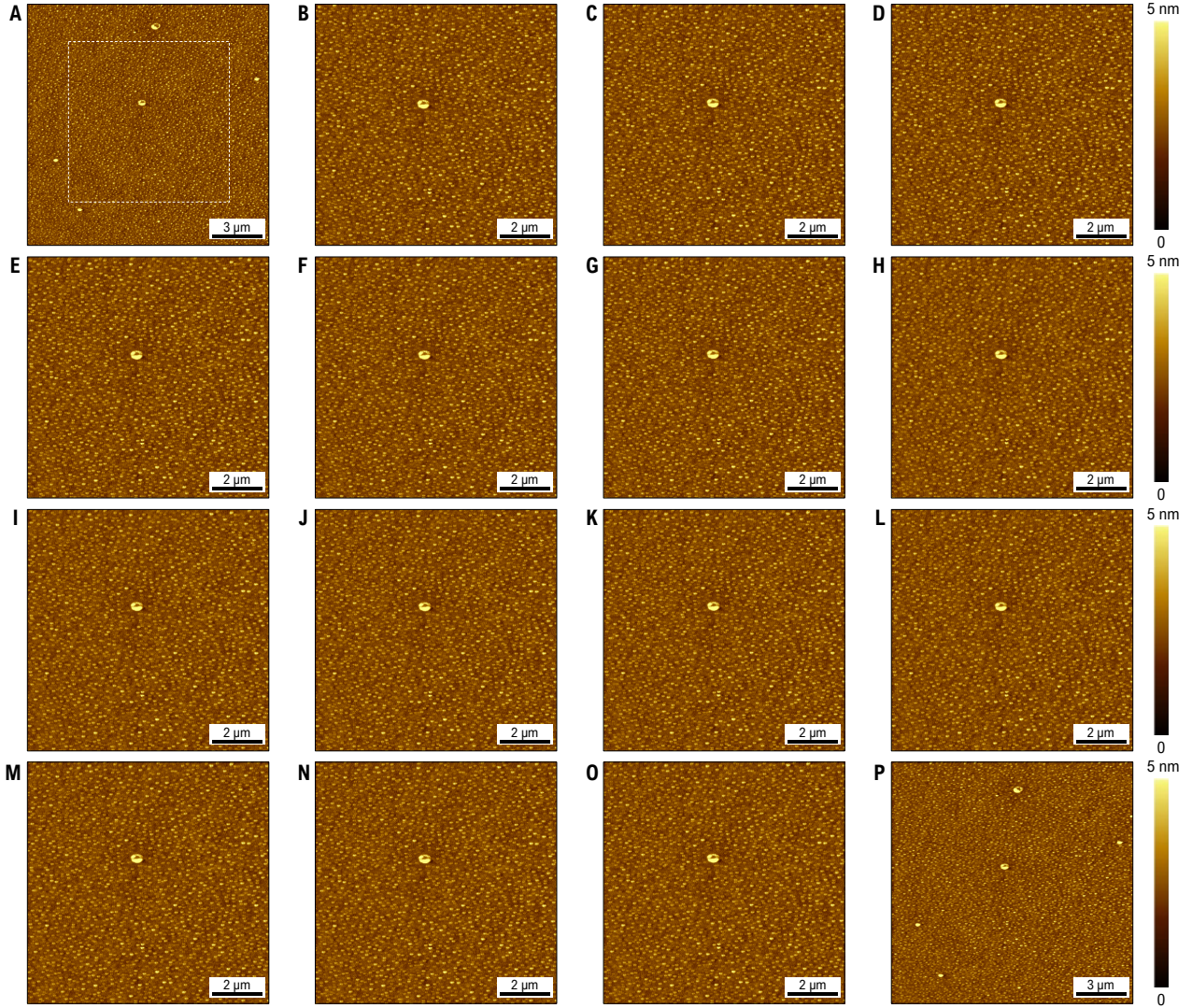

**Fig. S7. Non-destructive measurement and stability test of NC-HEsFM on PPLN.** (A) The initial topography image. (B-O) The topography images obtained by continuous *in-situ* scanning, and the scanning area is indicated by the white box in (A). (P) Final topography image measured after zoomed-in scanning. Measurement conditions: (A-C) $f_h$ = 521.685 kHz, $f_s$ = 554.218 kHz; (D, E) $f_h$ = 952.895 kHz, $f_s$ = 985.428 kHz; (F, G) $f_h$ = 1458.097, $f_s$ = 1490.630 kHz; (H, I) $f_h$ = 2034.985 kHz, $f_s$ = 2067.518 kHz; (J, K) $f_h$ = 2587.145 kHz, $f_s$ = 2619.678 kHz; (L, M) $f_h$ = 3239.841 kHz, $f_s$ = 3272.374 kHz; (N-P) $f_h$ = 4281.385 kHz, $f_s$ = 4313.918 kHz; (A-P) $f_{diff}$ = 32.533 kHz, $\Delta f_{sp}$ = +25 Hz, $A_{sp}$ = 10 mV.

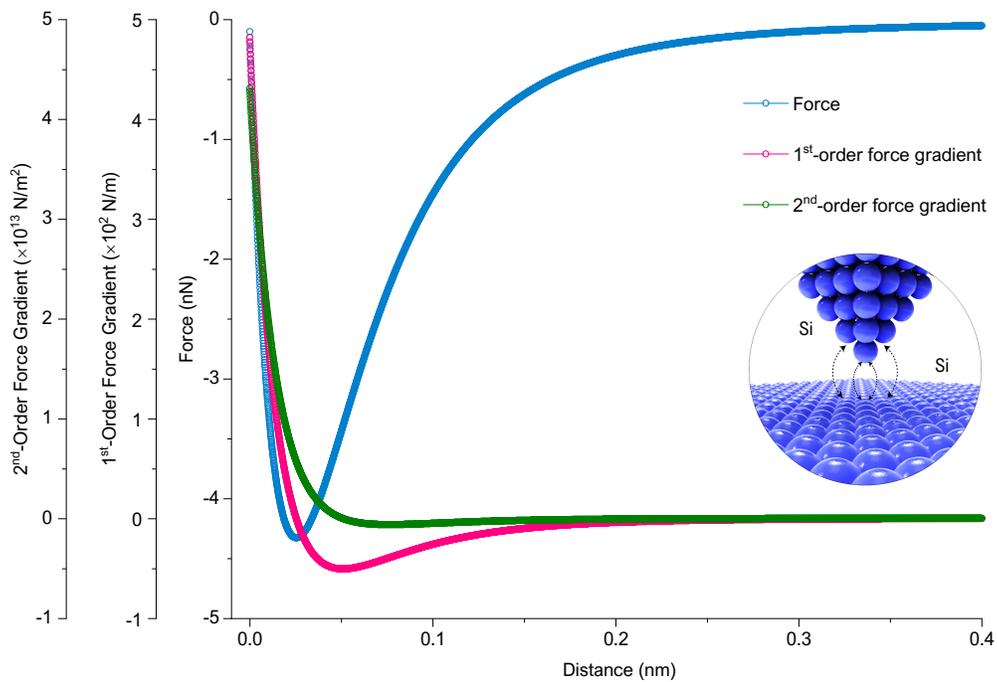

**Fig. S8. Tip-sample interaction force and force gradients as a function of tip-sample distance.**

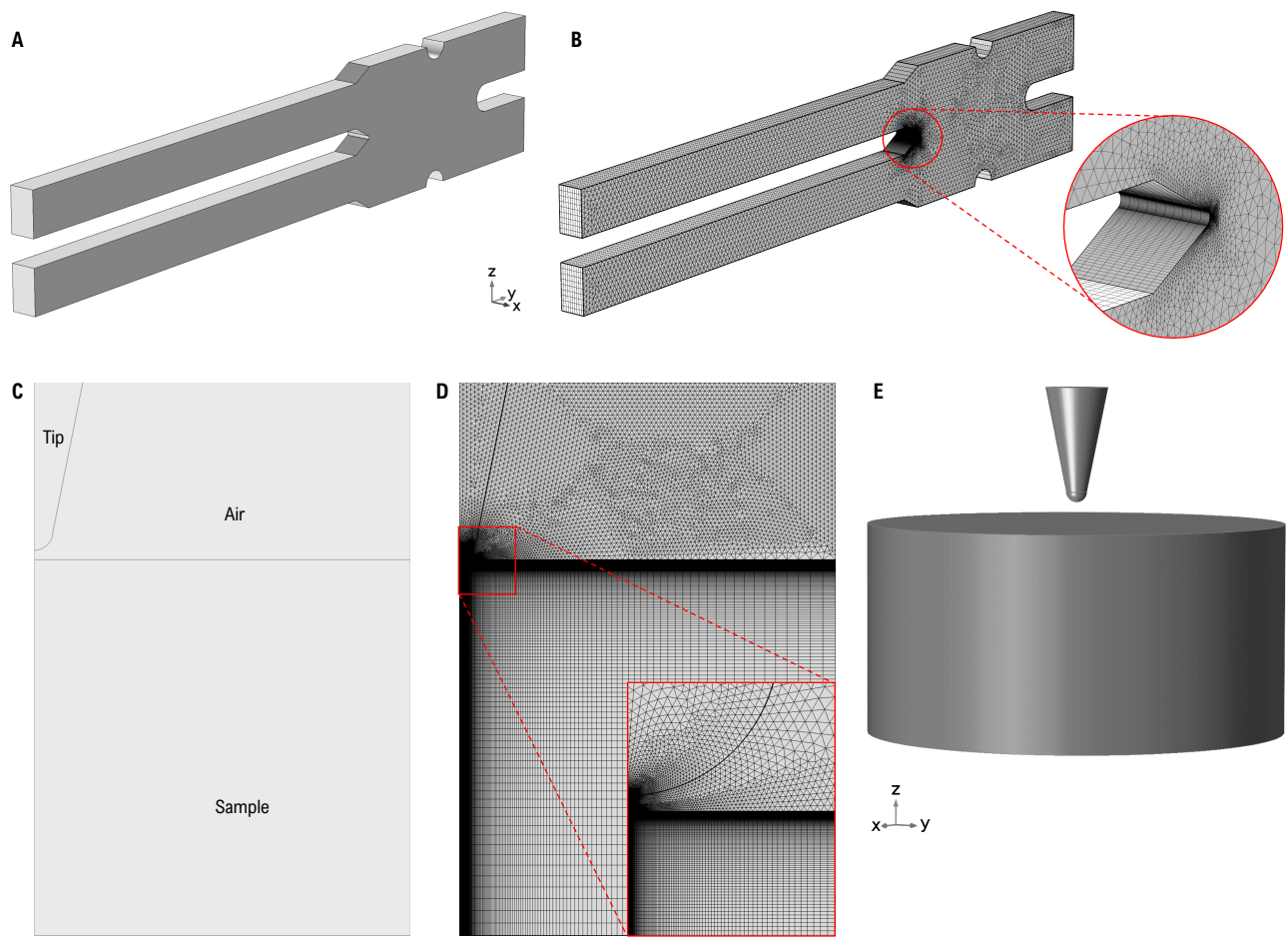

**Fig. S9. Models used in finite element simulations.** (A) 3D model and (B) the corresponding mesh used for QTF modal analysis. (C) 2D axisymmetric model and (D) the corresponding mesh used for the simulating the electric field distribution. (E) Schematic of the 3D data set generated from 2D revolution.

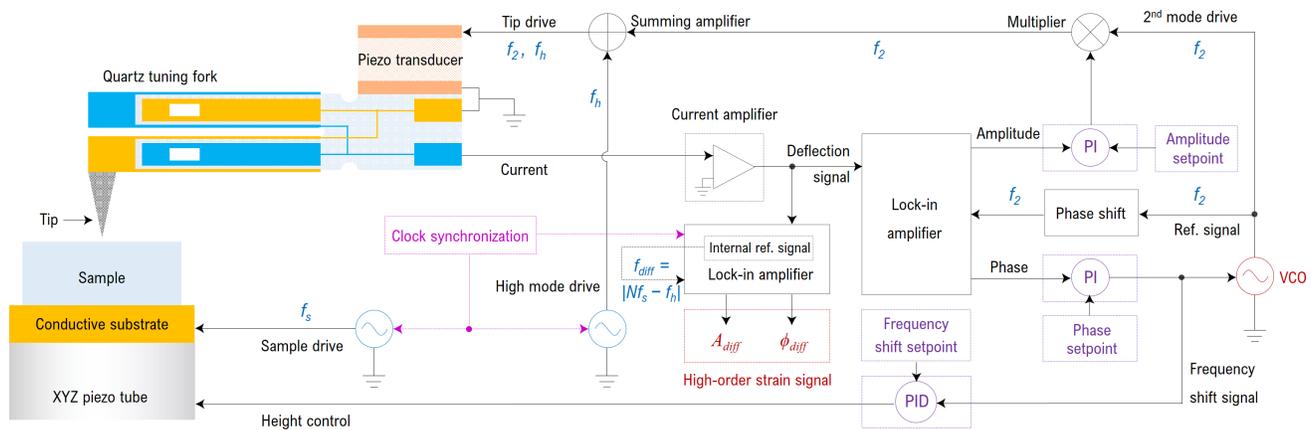

**Fig. S10. Schematic of the modified NC-HEsFM set-up for measuring high-order electrostrain.**

| QTF Mode No.      | 1      | 2       | 3      | 4      | 5       | 6       | 7       | 8       |
|-------------------|--------|---------|--------|--------|---------|---------|---------|---------|
| Simulation (kHz)  | 32.774 | 197.22  | 522.15 | 953.06 | 1456.8  | 2008.0  | 2582.9  | 3184.5  |
| QTF #1 (kHz)      | 32.674 | 197.439 | 521.68 | 952.89 | 1458.09 | 2014.98 | 2587.14 | 3209.84 |
| QTF #2 (kHz)      | 32.591 | 197.489 | 522.51 | 953.95 | 1459.56 | 2015.09 | 2594.12 | 3195.51 |

**Table S1. Simulated and experimentally measured eigenfrequencies of the QTF.**

# References

1. F. J. Giessibl, Forces and frequency shifts in atomic-resolution dynamic-force microscopy. *Phys. Rev. B* **56**, 16010-16015 (1997).

2. F. J. Giessibl, Atomic resolution on Si(111)-(7×7) by noncontact atomic force microscopy with a force sensor based on a quartz tuning fork. *Appl. Phys. Lett.* **76**, 1470-1472 (2000).

3. B. Babic, M. T. L. Hsu, M. B. Gray, M. Lu, J. Herrmann, Mechanical and electrical characterization of quartz tuning fork force sensors. *Sensors and Actuators A: Physical* **223**, 167-173 (2015).

4. O. E. Dagdeviren, U. D. Schwarz, Optimizing qPlus sensor assemblies for simultaneous scanning tunneling and noncontact atomic force microscopy operation based on finite element method analysis. *Beilstein J. Nanotechnol.* **8**, 657-666 (2017).

5. O. E. Dagdeviren, U. D. Schwarz, Numerical performance analysis of quartz tuning fork-based force sensors. *Meas. Sci. Technol.* **28**, 015102 (2016).

6. R. D. Grober *et al.*, Fundamental limits to force detection using quartz tuning forks. *Rev. Sci. Instrum.* **71**, 2776-2780 (2000).

7. M. V. Salapaka, H. S. Bergh, J. Lai, A. Majumdar, E. McFarland, Multi-mode noise analysis of cantilevers for scanning probe microscopy. *J. Appl. Phys.* **81**, 2480-2487 (1997).

8. P. R. Saulson, Thermal noise in mechanical experiments. *Phys. Rev. D* **42**, 2437-2445 (1990).

9. J. M. L. Miller *et al.*, Effective quality factor tuning mechanisms in micromechanical resonators. *Appl. Phys. Rev.* **5**, 041307 (2018).

10. Q. N. Chen, Y. Ou, F. Ma, J. Li, Mechanisms of electromechanical coupling in strain based scanning probe microscopy. *Appl. Phys. Lett.* **104**, 242907 (2014).

11. J. Yu *et al.*, Quadratic electromechanical strain in silicon investigated by scanning probe microscopy. *J. Appl. Phys.* **123**, 155104 (2018).

12. B. Chen *et al.*, Large electrostrictive response in lead halide perovskites. *Nat. Mater.* **17**, 1020-1026 (2018).

13. Y. Kim *et al.*, Nonlinear Phenomena in Multiferroic Nanocapacitors: Joule Heating and Electromechanical Effects. *ACS Nano* **5**, 9104-9112 (2011).

14. P. Girard, Electrostatic force microscopy: principles and some applications to semiconductors. *Nanotechnology* **12**, 485-490 (2001).

15. K. Zeng, Q. Zeng, Singapore Application No. 10202100483P (2021).